\documentclass[onecolumn]{emulateapj}

\begin{document}
\slugcomment{Submitted to ApJ: 2007/05/08; Accepted: 2007/06/15}
\shorttitle{Survey of $^{12}$CO/$^{13}$CO}
\shortauthors{Sheffer et al.}

\title{\textit{Hubble Space Telescope} Survey of Interstellar $^{12}$CO/$^{13}$CO in the Solar Neighborhood}
\author{Y. Sheffer\altaffilmark{1}, M. Rogers\altaffilmark{1,2}, S. R. Federman\altaffilmark{1}, D. L. Lambert\altaffilmark{3}, and R. Gredel\altaffilmark{4}}

\altaffiltext{1}{Department of Physics and Astronomy, University of Toledo, Toledo, OH 43606; ysheffe@utnet.utoledo.edu, steven.federman@utoledo.edu}
\altaffiltext{2}{Also at Marietta College, Marietta, OH 45750; rogerse@marietta.edu}
\altaffiltext{3}{Astronomy Department, University of Texas, Austin, TX 78712; dll@astro.as.utexas.edu}
\altaffiltext{4}{Max Planck Institut f\"{u}r Astronomie, K\"{o}nigstuhl 17, D-69117 Heidelberg, Germany; gredel@mpia.de}

\begin{abstract}
We examine 20 diffuse and translucent Galactic sight lines and extract the column densities of
the $^{12}$CO and $^{13}$CO isotopologues
from their ultraviolet $A$--$X$ absorption bands detected in archival Space Telescope Imaging
Spectrograph data with $\lambda$/$\Delta\lambda \geq$ 46,000.
Five more targets with Goddard High-Resolution Spectrograph data are added to the sample that more
than doubles the number of sight lines with published \textit{Hubble Space Telescope} observations
of $^{13}$CO.
Most sight lines have 12-to-13 isotopic ratios that are not significantly different
from the local value of 70 for $^{12}$C/$^{13}$C,
which is based on mm-wave observations of rotational lines in emission from CO and H$_2$CO
inside dense molecular clouds, as well as on results from optical measurements of CH$^+$.
Five of the 25 sight lines are found to be fractionated toward lower 12-to-13 values, while
three sight lines in the sample are fractionated toward higher ratios, signaling the predominance
of either isotopic charge exchange or selective photodissociation, respectively.
There are no obvious trends of the $^{12}$CO-to-$^{13}$CO ratio with physical conditions
such as gas temperature or density,
yet $^{12}$CO/$^{13}$CO does vary in a complicated manner with the column density of
either CO isotopologue, owing to varying levels of competition between isotopic charge exchange
and selective photodissociation in the fractionation of CO.
Finally, rotational temperatures of H$_2$ show that all sight
lines with detected amounts of $^{13}$CO pass through
gas that is on average colder by 20 K than the gas without $^{13}$CO.
This colder gas is also sampled by CN and C$_2$ molecules, the latter indicating gas kinetic
temperatures of only 28 K, enough to facilitate an efficient charge exchange reaction that lowers
the value of $^{12}$CO/$^{13}$CO.
\end{abstract}


\section{INTRODUCTION}

Measurements of isotopic ratios in the interstellar medium (ISM) are used to infer the past
history of chemical enrichment across the Galaxy.
Both quiescent stellar evolution as well as explosive stellar processes are sources of carbon
isotopes that enrich the ISM.
In our Solar System the carbon isotopic ratio, $^{12}$C/$^{13}$C, is found to be 89 \citep{ag89},
thus providing a benchmark value dating back 4.6 billion years to the time of
Solar formation.
Red giant stars are prodigious converters of $^{12}$C into $^{13}$C, and their ongoing mass
loss is a major factor in the continuous lowering of the 12-to-13 ratio in the ISM.
For example,
observations of CN, CO, and CS in highly-evolved AGB and post-AGB stars show $^{12}$C/$^{13}$C
values from 70 down to 30 in photospheric layers and in circumstellar shells that are being
injected into the ISM \citep{lambert86,kahane92}.
Planetary nebulae are later stages of low-mass stellar evolution, showing very low
values of $^{12}$CO/$^{13}$CO from 31 down to 2 \citep{balser02}, or from 23 down to 9 for the
sample in \citet{palla00}, as well as low 12-to-13 values from observations of \ion{C}{3}
\citep{clegg97,palla02,rubin04}.
Thus in the present-day, local ISM we should find $^{12}$C/$^{13}$C values that are between the
higher (older) Solar System value and the lowest (youngest) values found in highly-evolved objects.

The 12-to-13 carbon ratio should change not only in time, but also in space.
Specifically, it is expected to increase monotonically with Galactocentric distance
owing to lower cumulative numbers of star formation generations and $^{13}$C mass loss episodes
that have occurred farther from the Galactic center.
This picture has been confirmed by ISM studies employing the easily-detected
mm-wave emission lines of $^{12}$C- and $^{13}$C-bearing molecules to infer $^{12}$C/$^{13}$C
inside dense molecular clouds.
Observations of CO \citep{lp90} and H$_2$CO \citep{henkel82} show that the
ratio of the $^{12}$C-bearing molecules to their heavier counterparts are higher in the outer parts
of the Galactic disk than in its inner parts---see reviews by \citet{wr94} and \citet{wilson99},
as well as the work of \citet{savage02} and \citet{milam05}, who observed mm-wave
emission from CN in molecular clouds and derived a Galactic gradient similar to that seen in CO. 
However, appreciable scatter is found in the measured $^{12}$C/$^{13}$C at each Galactocentric
distance, which according to \citet{wilson99} is providing evidence for actual source-to-source
differences.
Averaging isotopic ratios from 13 sources in the local ISM gives 69 $\pm$ 6, while the fit
to the Galactic gradient predicts $^{12}$C/$^{13}$C = 71 $\pm$ 13 at the solar circle, indicating
some 25\% enrichment of $^{13}$C relative to $^{12}$C in the ISM over the last 4.6 eons. 

Another molecule used as a proxy for the $^{12}$C/$^{13}$C along diffuse and translucent sight
lines is CH$^+$, which has been investigated
primarily via its optical $A$--$X$ transition at 4232.548 \AA.
After a period of initial confusion, published $^{12}$CH$^+$/$^{13}$CH$^+$ values eventually
settled around an average value near 70 [e.g., \citet{crane91}, \citet{sw92}, \citet{hawkins93}],
confirming that CH$^+$ is not susceptible to fractionation because
its formation occurs via non-thermal processes.
Thus CH$^+$ provides confirmation that the \citet{wilson99} average of 69 $\pm$ 6 from
mm-wave emission data of dense molecular cloud cores is a good indicator of
$^{12}$C/$^{13}$C in the local ISM.
We shall, therefore, take the ambient value of $^{12}$C/$^{13}$C in the solar neighborhood to be
70 $\pm$ 7.

Besides mm-wave probing of dense clouds and CH$^+$ in much more tenuous lines of sight,
there are other cases that confirm the ambient carbon isotopic ratio of 70.
In the vacuum UV (VUV), \citet{sheffer02b} reported $^{12}$CO/$^{13}$CO = 73 $\pm$ 12 toward
X Per (HD 24534), albeit with highly fractionated C$^{16}$O/C$^{18}$O and C$^{16}$O/C$^{17}$O
values.
\citet{burgh07} reported 12-to-13 CO ratios for six sight lines, five of which resulted in
values that were consistent with 70 to within 2 $\sigma$.
Six out of seven $^{13}$CO sight lines in the sample of \citet{sonnentrucker07}
are consistent with unfractionated CO at the same level of significance.
In the near infrared (NIR), toward the dense cloud AFGL 490, \citet{goto03} derived
$^{12}$CO/$^{13}$CO = 86 $\pm$ 49, i.e., indistinguishable from the ambient carbon ratio.
The sight line toward the massive protostar NGC 7538 IRS9 provided the first isotopic
measurement of solid CO in the ISM.
NIR data presented in \citet{boogert02} show that the ratio of $^{12}$CO ice to $^{13}$CO ice
is 71 $\pm$ 5.
Moreover, this agreement is also displayed by ices of another C-bearing molecule,
since $^{12}$CO$_2$/$^{13}$CO$_2$ = 80 $\pm$ 11 along the same line of sight \citep{boogert00}.

As the examples above show, most isotopic carbon ratios are inferred from C-bearing molecular
proxies, especially CO, thanks to the larger isotopic shifts that are encountered in molecules.
However, in the ISM, carbon isotopic ratios that are derived via molecular proxies are
sometimes found to be fractionated.
Two opposing effects are usually invoked to explain CO fractionation away from the ambient
$^{12}$C/$^{13}$C.
An isotopic charge exchange (ICE) reaction $^{13}$C$^+$ + $^{12}$CO $\longrightarrow$ $^{12}$C$^+$
+ $^{13}$CO + 35 K \citep{watson76} works to lower the
$^{12}$CO-to-$^{13}$CO ratio when a lower gas temperature and a high C$^+$ abundance are present,
while selective photodissociation (SPD) in ultraviolet (UV) photon-dominated regions drives the
ratio higher
through self shielding by the more abundant isotopologue, i.e., $^{12}$CO \citep{bl82,vdb88}.
Deep inside dark molecular clouds no fractionation of CO is expected because of the absence of both
ionized carbon and UV photons.

In order to explore how well $^{12}$CO/$^{13}$CO in diffuse molecular clouds reflects
the underlying 12-to-13 carbon ratio, we study its fractionation processes here in more detail
through a larger sample of sight lines.
It is important to verify that the 12-to-13 variations are source-related and not induced
by systematic effects such as noise and modeling uncertainties.
Therefore, because the treatment of absorption along an infinitesimal beam is significantly less
complicated than that of emission from a finite volume, and since the
Space Telescope Imaging Spectrograph (STIS) records
high-quality spectra of a large number of saturated as well as optically-thin
CO bands per sight line, we are able to provide evidence for source-to-source variation
in $^{12}$CO/$^{13}$CO, and consider photochemical fractionation as its cause. 

\section{DATA, REDUCTION, AND SPECTRUM SYNTHESIS}

Our sample includes 20 stars observed with STIS at resolving power ($R$) between 46,000 (grating
E140M) and 160,000 (grating E140H) and five stars that have \textit{Hubble Space Telescope (HST)}
data from its Goddard High-Resolution Spectrograph (GHRS).
Although $R$ of the GHRS grating G160M is the lowest in the current sample (19,000),
with proper sampling of weak and strong bands and with high enough signal-to-noise ratio (S/N),
the data are very useful for robust modeling of CO column densities.
This was nicely demonstrated by \citet{lambert94} who presented G160M data for $\zeta$ Oph
having S/N $\sim$ 900(!) and by \citet{federman03}, who modeled CO toward $\rho$ Oph A and
$\chi$ Oph using G160M data with S/N $\sim$ 200.
These three stars were added to the sample, as well as HD 96675 and HD 154368, which have
GHRS spectra with only moderate S/N;
the results of their low-$R$ CO modeling are less secure than for the rest of the sample.

Table 1 lists all 25 sight lines and gives stellar spectral
types, apparent $V$ magnitudes, Galactic coordinates, velocity corrections to the Local Standard
of Rest (LSR), reddening values, and heliocentric and Galactocentric distances. 
The farthest target examined here is HD 177989, which lies 5 kpc away.
Table 2 provides a listing of STIS and GHRS data sets that were used in our analysis.
Also listed for each data set are grating and aperture used, S/N achieved,
and $R$, derived as a fitted parameter during CO spectral syntheses.
Whenever repeated multiple exposures were available, they were combined in wavelength space for
improved S/N.
In addition, when a feature (band) appeared in two adjacent orders, they were combined after
correcting for any small wavelength inconsistencies.
All reductions here were performed in IRAF\footnotemark[1] and STSDAS.\footnotemark[2]

\footnotetext[1]{IRAF is distributed by the National Optical Astronomical Observatory, which is
operated by the Association of Universities for Research in Astronomy, Inc., under cooperative
agreement with the National Science Foundation.}
\footnotetext[2]{STSDAS is a product of the Space Telescope Science Institute, which operated
by AURA for NASA.}

All profile syntheses of CO, as well as of H$_2$ from spectra acquired with the
\textit{Far Ultraviolet Spectroscopy Explorer (FUSE)}, were carried out by Y. S. with his simplex
Fortran code, Ismod.f.
For the determination of CO column density ($N$) we used oscillator strengths ($f$-values) for the
$A$--$X$ bands from \citet{chan93}, see also \citet{mn94}, which in a global sense
have been verified to a level of a few percent by \citet{eidelsberg99}.
Simultaneous multi-parametric profile syntheses of all detected CO bands for each sight line were
employed in deriving $N$($^{12}$CO), $N$($^{13}$CO), and excitation temperatures ($T_{\rm ex}$),
as well as observed radial velocity and fitted equivalent width ($W_{\lambda}$) for each band.
Cloud component structures from high-$R$ optical observations of CH $\lambda$4300 were available
for 19 of the 22 sight lines analyzed here, but
relative fractions and $b$-values of all components were allowed to vary (i.e., only
velocity differences were left fixed during the fit).
The remaining three cases without available CH component structures happen to be the sight lines
with the lowest amounts of detected CO, thus affording uncomplicated abundance determinations.
Detailed cloud structures and model results for CH and H$_2$ will appear elsewhere.
Here, values for $N_{\rm tot}$ of H$_2$ and its excitation temperatures $T_{\rm 1,0}$ and
$T_{\rm 4,0}$ will be quoted and analyzed.

The important aspect of correctly handling any effects of line saturation was realized by including
both very weak $A$--$X$ bands with optical depth at line center of $\sim$ 1 or less, which are
sensitive mostly to the total CO column density, as well as very strong $A$--$X$ bands having
$f$-values many times larger, which dictate how the cloud components should be modeled.
Figures 1 and 2 show two examples of multi-band fits toward HD 208266 of $^{12}$CO and
$^{13}$CO, respectively.
The wide spectral coverage and the high abundance of CO along this sight line combine to give us
the highest practical number (14) of usable absorption bands for $^{12}$CO, as well as seven
consecutive bands of $^{13}$CO.
In terms of optical depth ($\tau$), the model of $^{12}$CO shows that $\tau$ increases by a factor
of 1150 from $A$--$X$ (13--0) to $A$--$X$ (2--0), whereas $\tau$ of the $^{13}$CO bands varies by
a factor of 20.
It should be noted that a full-scale fit of 14 $A$--$X$ bands involves simultaneous modeling of
the column density in 126 VUV transitions of CO, when the model includes three parametrized
excitation temperatures, or nine rotational transitions, per band.
This large number of modeled lines is what makes such a global fit very robust, even when the data
(from E140M) have moderate $R$ and S/N.

Some details from these spectral fits, as well as model $N$ values, are given in Table 3 for
$^{12}$CO and in Table 4 for $^{13}$CO.
Results are listed for the two $A$--$X$ bands (denoted by
$v^\prime$) that straddle the $\tau$ = 1 condition at the center of the R(0) line.
These results include fitted $W_{\lambda}$, optical depth, and column density.
The two bands are then used to infer the uncertainty on the derived column density by averaging
their relative uncertainties in equivalent width and by assuming that the result is identical
to the relative error in the column density.
Of course, the lower the optical depth, the better is this assumption.
However, one cannot use absorption bands that are too weak ($\tau \ll$ 1), because of the increasing
influence of the noise relative to the decreasing signal from CO.

The modeling methodology was to derive a cloud structure for the $^{12}$CO molecule, and
then to apply it as a fixed structure during modeling of $^{13}$CO.
The number of components, their relative fractions, and their velocity separations were all kept
fixed, under the assumption that the two isotopologues reside in the same clumps of gas, with all
components having identical 12-to-13 ratios for a given sight line.
This assumption is kinematicaly verified, since the radial velocities of both species in our
sample agree very well with each other.
This agreement is shown in Fig. 3, a plot of the difference $V_{\rm helio}(^{13}{\rm CO}) -
V_{\rm helio}(^{12}{\rm CO})$ against the heliocentric radial velocity scale of $^{12}$CO.
With an average of $-$0.2 $\pm$ 0.4 km s$^{-1}$, the scatter in this sample is consistent
with a vanishing radial velocity difference between the two isotopologues.
The 1 $\sigma$ velocity dispersion of 0.4 km s$^{-1}$ is a small fraction of the resolution
element for the data,
being 0.22$\Delta\lambda$ and 0.06$\Delta\lambda$ for E140H and E140M, respectively.

Although it is known that the STIS line spread function (LSF) possesses weak wings (see the STIS
handbook), we could not see any obvious signatures for such wings in the highest-$R$ data.
Therefore, a single Gaussian was used to describe the instrumental profile of STIS.
This is is agreement with the findings of \citet{jt02}, who analyzed the unbinned (but
noisier and lacking proper flat-fielding) raw spectral data.
Nevertheless, the STIS handbook does show that the LSF wings have a larger role in the
profile for wider slits.
Therefore, $R$ was treated as a fittable parameter during modeling, and
it is apparent that $R$ is lower for E140H data taken through wider slits (see Fig. 4).
There is a good match between the range of $R$ found here and
the range found by \citet{bowers98} for the 0$\farcs$09 slit, as well as his listed
range for the smallest (``Jenkins'') slit based on ground (pre-launch) measurements.
Furthermore, our tests show that a 15\% error in $R$ affects the derived
column density by only $\sim$ 6\%. Since both $N$($^{12}$CO) and $N$($^{13}$CO) vary similarly
in response to any $R$ change, derived isotopic ratios are affected at a level
of $\lesssim$ 3\% according to these tests.

The trend of lower $R$ with larger slit width is not limited to the E140H grating, as shown by
modeling medium-$R$ data from the E140M grating.
Three sight lines return an average of $R$ = 45,600 $\pm$ 1,000 for the 0$\farcs$06 slit,
in excellent agreement with the value 46,000 given in \citet{bowers98}. On the other hand,
a couple of sight lines were observed through the wider 0$\farcs$2 slit, and their fitted
$R$ values are 38,000 and 39,000, i.e., clearly corroborating the trend of decreasing $R$
with larger slit widths.

\section{COMPARISON WITH PREVIOUS RESULTS}

\subsection{\it Internal Comparison with Previous Ismod.f Results}

The results presented in this paper reflect simultaneous modeling of all detectable $A$--$X$ bands
per sight line, but initial Ismod.f fits of certain sight lines did not use all available bands.
It is of interest to compare the latest and most extensive fits with less ambitious initial
modeling to learn about the dependence of model results on the number
of bands being fitted simultaneously.
\citet{sheffer02a} modeled a four-component cloud structure of $N$($^{12}$CO) and $N$($^{13}$CO)
toward HD 24534 using the
strongest and weakest of only seven $A$--$X$ bands that were available from the STIS o648o12--13
data sets, as well as any available intersystem bands.
For comparison and completeness, we obtained and analyzed the STIS o66p01--2 data sets from the
\textit{HST} archive and modeled a total of 12 $A$--$X$ bands toward X Per, using all four E140H
data sets.
Keeping a four-component cloud model, the fit returned $N$($^{12}$CO) = (1.58 $\pm$ 0.04) $\times$
10$^{16}$ cm$^{-2}$, which is 12\% higher than the value in \citet{sheffer02a}.

However, re-synthesizing the $^{13}$CO bands toward X Per returned a new column density of 1.55
$\times$ 10$^{14}$ cm$^{-2}$, which was lower by 20\% than the \citet{sheffer02a} value. 
This newer fit was based on 9 bands covering a robust range of optical depth and $f$-values.
A closer inspection revealed that the weaker bands were not being fitted correctly,
showing as excess of observed absorption relative to the model.
We, therefore, relaxed the usual method of keeping the component structure that was derived from
$^{12}$CO fixed during $^{13}$CO fits, by allowing the $b$-values to vary during the synthesis.
This resulted in a better fit to the data, yielding a larger $N$($^{13}$CO), albeit 4\% below
the value in \citet{sheffer02a}.
Thus our published value (73 $\pm$ 12) for $^{12}$CO/$^{13}$CO is updated here to 85 $\pm$ 5,
or higher by 1.0 $\sigma$.
Since more bands were used in the latest fit, its results should be more secure than before.
The combination of high CO column density, $N$($^{12}$CO) $\geq$ 10$^{16}$ cm$^{-2}$ and
$N$($^{13}$CO) $>$ 10$^{14}$ cm$^{-2}$, and high S/N in observed low-$v^\prime$ (stronger) bands of
$^{13}$CO, is what allowed us to detect the anomaly in the first place and then to treat it.
Both $^{12}$CO and $^{13}$CO are still assumed to reside in the same clumps of gas, albeit with
different line widths.
Another option, that of allowing a fit of cloud component fractions, would have explored a new
territory of variable isotopic ratios among clumps of gas, which we decided not to pursue
at the present time.

Toward HD 203374A, we previously published two $N$($^{12}$CO) values based on the five $A$--$X$
bands (7--0) through (11--0) from the E140H data set, together with the $C$--$X$ (0--0) band from
lower-$R$ \textit{FUSE} data, which provided the high-$\tau$ leverage for the Ismod.f synthesis
\citep{sheffer03,pan05}.
Here, the much more extensive set of bands available in the E140M (lower-$R$) data set
was analyzed.
A simultaneous fit of 112 transitions in 13 $A$--$X$ bands, (0--0) through (12--0),
returned a value for $N$($^{12}$CO) only 6\% higher than \citet{sheffer03}, or a value
that agrees perfectly with \citet{pan05}.
This good agreement demonstrates that one need not resort to the highest-$R$ available
(e.g, through E140H) as long as
there is an adequate and robust coverage of many bands from medium-$R$ (E140M) exposures.

\subsection{\it Comparison with Previous External Results}

Since relatively little has been done concerning UV data on $^{13}$CO in the past, most of the
common sight lines involve comparisons of $^{12}$CO results (see Table 5). 
Previously published $^{13}$CO sight lines with \textit{HST} observations include $\zeta$ Oph
\citep{sheffer92,lambert94}, X Per \citep{sheffer02a}, and $\rho$ Oph~A and
$\chi$ Oph \citep{federman03}, all modeled with Ismod.f.
Two recent studies by \citet{burgh07} and \citet{sonnentrucker07} investigated $^{12}$CO,
$^{13}$CO, and H$_2$ along diffuse and translucent sight lines.
Lines of sight with detections of $^{13}$CO in the two studies amounted to 6/23 and 7/10 of the
whole samples, respectively, of which four targets were common to both.
Together, these two studies introduced six sight lines with previously unpublished \textit{HST}
data for $^{13}$CO: HD 27778, HD 177989, HD 203532, HD 206267, HD 207198, HD 210121, and HD 210839.
The current sample presents independent data reduction and analysis of 22 $^{13}$CO sight lines,
of which 15 have no previously published $^{12}$CO-to-$^{13}$CO results, bringing the total number
of sight lines with detectable $^{13}$CO based on \textit{HST} data from 11 to 26.

\citet{burgh07} modeled their data with a single cloud component, resulting in a single
``effective'' $b$-value for all their sight lines.
Furthermore, instead of using a self-fitting code, \citet{burgh07} created a grid of models and then
chose the values of $N$, $T_{\rm rot}$, and $b$ that best described the data.
On the other hand, \citet{sonnentrucker07} incorporated a profile-fitting code with multi-component
cloud structures from
high-$R$ optical spectra of CH and CN, much like our usual method of analysis.
We are convinced that the latter methodology results in more realistic characterization
of each line of sight, as is evidenced from the overall larger uncertainties associated
with the results of \citet{burgh07}.
Both \citet{burgh07} and \citet{sonnentrucker07} employed the CO $f$-values listed in \citet{mn94},
which are the values from \citet{chan93}.
Thus no re-scaling was needed for this comparison with our results, which are also based on the
same set of $f$-values.
Only a single star from \citet{sonnentrucker07} (HD 210121) is not included in the current sample
because there is a very limited coverage by GHRS exposures.
By not including literature results our entire sample is kept
as homogeneous as possible in terms of data reduction, measurements,
and Ismod.f spectrum synthesis fits.
Table 5 shows that $^{12}$CO/$^{13}$CO results for the seven common sight lines are in agreement
to within 1.5 $\sigma$, with the sole
exception of the \citet{kaczmarczyk00} value for X Per owing to improper input
cloud structure during modeling \citep{sheffer02a}.

\section{ISOTOPIC RESULTS AND ANALYSIS}

\subsection{\it $^{12}$CO/$^{13}$CO Sample}

Figure 5 shows log $N$($^{13}$CO) versus log $N$($^{12}$CO) for all sight lines.
The slope corresponding to $^{12}$C/$^{13}$C = 70 is shown, together with its 1 $\sigma$ range,
as well as the slopes of 0.5 and 2.0 times the ratio of 70.
The values of $^{12}$CO/$^{13}$CO are seen to be centered along the line
$^{12}$C/$^{13}$C = 70, showing considerable scatter but mostly within the factor-of-2
range relative to 70, i.e., within 35 to 140.

Another view is provided by Fig. 6, which shows a histogram of the current VUV $^{12}$CO/$^{13}$CO
values, as well as distributions via other detection methods and of other C-bearing molecules.
This VUV sample of CO ratios is at the bottom, showing a distribution that is centered near the
ambient carbon isotopic ratio (70) with roughly equal
numbers of sight lines having lower or higher values.
Table 6 shows that $^{12}$CO/$^{13}$CO in diffuse molecular clouds varies between
37 $\pm$ 8 and 167 $\pm$ 25, a factor of 4.5 $\pm$ 1.2.
The almost symmetric distribution is betrayed by 11 sight lines with 12-to-13 ratios
below 70 and by 14 sight lines with values above 70.
Six UV sight lines, all common with this sample, were analyzed by \citet{burgh07} and yielded
$^{12}$CO/$^{13}$CO values over a narrow range of 49 $\pm$ 15 to 68 $\pm$ 31.
For a more proper comparison, the same six targets here returned values between
41 $\pm$ 7 and 85 $\pm$ 5, in very good agreement with \citet{burgh07}.
However, the sample of \citet{burgh07} is dominated by larger 12-to-13 uncertainties that
obscure any possible source-to-source variations, whereas here we are able to discern
source-to-source variations thanks to smaller measurement and modeling uncertainties.
Another sample of six UV sight lines from \textit{HST} data in \citet{sonnentrucker07} shows
CO isotopic values between 46 $\pm$ 6 and 79 $\pm$ 12 for five sight lines common with this sample,
from which the range is between 42 $\pm$9 and 85 $\pm$ 5, revealing very good agreement as well.

Table 6 also lists the isotopic ratios normalized by the carbon ratio, or
\begin{displaymath}
F_{13} \equiv \frac{^{12}{\rm CO}/^{13}{\rm CO}}{^{12}{\rm C}/^{13}{\rm C}} = \frac{^{12}{\rm CO}/^{13}{\rm CO}}{70}.
\end{displaymath}
In terms of $F_{13}$, CO isotopic values are found between 0.5 and 2.4, yet very well centered
on the ambient carbon isotopic value, or $F_{13}$ = 1.
This distribution is consistent with $^{12}$C/$^{13}$C being the ``parental''
distribution for the CO isotopic ratios.
Theoretical models \citep{vdb88} predict ambient isotopic CO ratios for diffuse
clouds, and variable ratios that respond to the changing UV flux along a path into a more
opaque cloud.
Yet the same models fail to predict the observed column densities of CO
along diffuse sight lines.
Better correspondence may occur once new measurements of $f$-values
of Rydberg transitions of CO ($\S$ 4.4) are incorporated by such modeling efforts.

\subsection{\it Sight Lines with Fractionated CO}

An examination of CO isotopic deviations from 70 shows that 14/25 of sight
lines agree with the value of ambient $^{12}$C/$^{13}$C at the level of $\leq$ 1 $\sigma$, and that
three more sight lines agree with the unfractionated value to within 2 $\sigma$.
Thus 32\% of sight lines have values of $^{12}$CO/$^{13}$CO that are significantly
($\geq$ 3 $\sigma$) different from the carbon isotopic value.
Out of these eight sight lines, three have CO that is fractionated upward: X Per = 85 $\pm$ 5,
$\zeta$ Per = 108 $\pm$ 5, and $\zeta$ Oph = 167 $\pm$ 25 (in order of increasing $F_{13}$),
presumably owing to more effective self shielding of $^{12}$CO against photodissociation.
The five other sight lines in order of decreasing $F_{13}$ are: HD 207538 = 51 $\pm$ 6, HD 207198
= 48 $\pm$ 8, HD 206267 = 42 $\pm$ 9, HD 203532 = 41 $\pm$ 7, and HD 154368 = 37 $\pm$ 8.
These exhibit fractionation in $^{13}$CO, presumably owing to a larger role of the ICE
reaction with C$^+$ that enhances the relative abundance of $^{13}$CO. 
\citet{sonnentrucker07} also found CO fractionated toward HD 206267, with $^{12}$CO/$^{13}$CO
= 46 $\pm$ 6.

The first 12-to-13 VUV study of CO based solely on \textit{HST} data was provided by
\citet{lambert94}, who found $^{12}$CO/$^{13}$CO to be 167 $\pm$ 25 toward $\zeta$ Oph.
[Previously, \citet{sheffer92} derived a 12-to-13 ratio of 150 $\pm$ 27 along the same sight
line using GHRS data and IUE results.]
Furthermore, two more sight lines in Ophiuchus did indicate high UV ratios as reported by
\citet{federman03}, who found 12-to-13 ratios of 125 $\pm$ 23 toward $\rho$ Oph A and
117 $\pm$ 35 toward $\chi$ Oph.
When such initial UV reports of highly fractionated CO came out they seemed to be extraordinary.
Now the present distribution serves to show that nothing extraordinary is happening even
along the line of sight toward $\zeta$ Oph.
Although it still shows the highest $^{12}$CO
fractionation (factor of 2.4 $\pm$ 0.4), another sight line, that of HD 154368, has $^{13}$CO
fractionated to the greatest extent of 1.9 $\pm$ 0.4, while six other sight lines were
listed above as fractionated to lesser degrees.
Thus UV data toward diffuse/translucent sight lines is showing that relative to
the local isotopic ratio of carbon, CO can be fractionated upward or downward by a factor of
$\lesssim$ 2.
For translucent sight lines, \citet{vdb88} models can easily duplicate downward,
but not upward, fractionations by about 2, showing the need to improve upon self shielding
computations for $^{12}$CO.

We now explore possible relationships between the observed values of $^{12}$CO/$^{13}$CO and the
other observables that characterize the ISM.
This may provide insight into the required combination of parameters that control the observed
distribution of the CO isotopic ratio.

\subsection{\it Variation of $^{12}$CO/$^{13}$CO with CO Column Density}

Figure 7 (filled circles) shows the 12-to-13 CO isotopic ratio versus $N$($^{12}$CO).
The global sense of the plot shows variations in the values of $^{12}$CO/$^{13}$CO for
different $N$($^{12}$CO) regimes. [Similar variations are seen when $^{12}$CO/$^{13}$CO
is plotted against $N$($^{13}$CO).]
Higher values of the isotopic ratio (above 70) are seen for the lowest column densities,
or for the regime of $N$($^{12}$CO) $\lesssim$ 2 $\times$ 10$^{15}$ cm$^{-2}$.
The second regime is found for $N$($^{12}$CO) $\lesssim$ 6 $\times$ 10$^{15}$ cm$^{-2}$,
where a significant number of sight lines is found with low 12-to-13 values down to about 50,
mixed with sight lines that still show isotopic values higher than 70.
Both of these regimes are influenced by UV-dominated photochemistry.
The higher values correspond to enhanced photodissociation of $^{13}$CO.
$^{12}$CO is already optically thick enough to be self-shielded against UV
photodissociation, since according to the self-shielding models of \citet{vdb88}, the
photodissociation rate of CO is diminished by a factor of $\approx$ $e^{-1}$ once the column
density of $^{12}$CO has reached the value of $\approx$ 10$^{15}$ cm$^{-2}$.
Moreover, taking into account the extra shielding from H$_2$ spectral features, the shielding
of $^{12}$CO commences at lower values, $N$ $\leq$ 10$^{14.5}$ cm$^{-2}$.
The regime that includes low $^{12}$CO/$^{13}$CO values probably corresponds to the ICE
reaction with C$^+$, which works to increase the abundance of $^{13}$CO at the
expense of $^{12}$CO \citep{watson76}.
Here, UV radiation is needed to ionize C atoms.
Finally, at sufficiently high column densities of $^{12}$CO, the third regime in the plot
shows mostly unfractionated isotopic ratios, i.e., values that are
consistent with $^{12}$C/$^{13}$C = 70.
In this regime, $^{13}$CO has become mutually shielded by $^{12}$CO \citep{vdb88}.

Also shown in Fig. 7 (empty circles) are $^{12}$CO/$^{13}$CO determination from \citet{ll98b},
based on mm-wave absorption observations, which show some overlap with VUV results in the middle
regime described above, i.e., where low 12-to-13 values are thought to be strongly influenced by
ICE.
But toward the third regime of highest $N$($^{12}$CO) values, the mm-wave absorption
results are significantly lower than VUV results, signaling that the two are drawn from different
populations of sight lines, with the radio observations probably probing much colder gas.

Qualitatively, there is an interesting resemblance between this plot and the solid line in Fig. 11
of \citet{vdb88}, which models the variation of $^{13}$CO/$^{12}$CO
as function of depth into a cloud, provided the solid line is inverted to reflect
$^{12}$CO/$^{13}$CO variations instead.
Figure 7 may be interpreted the same way, since $N$($^{12}$CO) of the abscissa registers the
amount of material along the line of sight (increasing optical depth), which can also stand for
the amount of molecular material along a path into a cloud.

\subsection{\it Fractionated Values and Their Genesis}

We surmise that all CO isotopic ratios would equal $\sim$ 70, were it not for ICE
or SPD pushing the intrinsic $^{12}$C/$^{13}$C toward lower or higher values.
A dedicated code is required for a full treatment of photodissociation rates and number densities
as influenced by a given UV radiation field and chemical abundances \citep[e.g.,][]{vdb88,
warin96}.
Instead, we compare our determinations
to values and results from \citet{vdb88} using a more simplified approach.
In analyzing $^{12}$CO/$^{13}$CO toward $\zeta$ Oph, \citet{lambert94}
introduced a formula that allows an evaluation of the isotopic ratio in terms of the
competition between SPD and ICE, namely,
$$F_{13} = \frac{\Gamma_{13} + n(^{12}{\rm C}^+)k_1^r}{\Gamma_{12} + n(^{12}{\rm C}^+)k_1^f},$$
where $\Gamma_i$ stands for the photodissociation rate of the isotopologue $^i$CO, $k_1^f$ is
the forward rate coefficient for the ICE reaction with C$^+$, the number density of which is
designated by $n$.
Without ICE, $F_{13}$ is simply the ratio of $\Gamma_{13}$ to $\Gamma_{12}$, which for
$N$($^{12}$CO) = 10$^{14}$ cm$^{-2}$ is only 1.2, but is $\approx$ 4.5 for $N$($^{12}$CO) =
10$^{16}$ cm$^{-2}$ \citep[][their Table 5]{vdb88}.
When ICE is the only process operating, the equilibrium value for $F_{13}$ is $k_1^r$/$k_1^f$
= exp($-$35/$T_{\rm kin}$), i.e., it depends only on the kinetic temperature of the gas.
It will be seen in $\S$ 4.8 that the average $T_{\rm kin}$ $\approx$ 60 K, which leads to $F_{13}$
$\approx$ 0.56,
or a 12-to-13 CO ratio of $\approx$ 39, under the operation of ICE without SPD.
Such a ratio is remarkably similar to the lowest values found in the current survey.

The formula for the photodissociation rate,
$$\Gamma_i = 2.04 \times 10^{-10} I_{\rm UV} \Theta_i \theta_c \quad{\rm s}^{-1},$$ 
includes the enhancement in the strength of the UV radiation field, $I_{\rm UV}$,
the shielding function of the isotopologue $^i$CO,
$\Theta_i$, and the continuum extinction owing to dust, $\theta_c$ \citep{vdb88}.
The values for $\Theta_i$ are extracted via interpolation from Table 5 of \citet{vdb88},
using one-half of the measured values for $N$(H$_2$) and $N$(CO) as input,
as appropriate for a finite modeled slab with equal UV illumination on both sides.
Other cloud geometries and/or fragmentation have been shown to significantly affect the CO
abundance relative to H and H$_2$, but not to affect the 12-to-13 CO isotopic ratio \citep{kopp00}.
The continuum dust extinction is parametrized
by exp($-\tau_{\rm UV}$) for the sake of consistency with the chemical modeling ($\S$ 4.5).
The number density of ionized carbon, $n(^{12}{\rm C}^+)$, is computed from
$1.4\times10^{-4}n(\rm H)$, the observed abundance of C$^+$ in the ISM \citep{cardelli96}.
The total gas (hydrogen) number density, $n_{\rm tot}$(H), is given by $n_{\rm CN}$, which in turn
is also determined from chemical modeling of $N$(CN) and $N$(CH) in $\S$ 4.5.
The values of $T_{\rm kin}$, $I_{\rm UV}$, $\tau_{\rm UV}$, $n_{\rm CN}$, $\Theta_{12}$, and
$\Theta_{12}$ are given for each sight line in Table 6.

Figure 8 shows observed $F_{13}$ values (empty squares) versus log $N$($^{12}$CO).
Using all the fractionation parameters listed in Table 6, we computed $\Gamma_{12}$ and
$\Gamma_{13}$, and then $F_{13}$ values for all sight lines, which are denoted by the `X'
symbols in Fig. 8.
The best correspondence between the observed and predicted values was found by minimizing the rms
differences between the two sets of $F_{13}$ values for all sight lines.
Such rms minimization was done in order to determine the value of any parameter that
best reproduces observed $F_{13}$ values.
When both fractionation processes are operating the predicted $F_{13}$ values
cannot easily reach above $\approx$ 1.0, unless values that are $\approx$ 5 times higher than
those found by the chemical models are incorporated into $I_{\rm UV}$, or values that are
$\approx$ 5 times lower are incorporated into $n(C^+)$ and/or $k_1^f$.
Two values have been given in the literature for $k_1^f$: $2 \times 10^{-10}$ cm$^3$ s$^{-1}$
\citep{watson76} and $7.5 \times 10^{-10}$ cm$^3$ s$^{-1}$ \citep{sa80}.
The smaller value for $k_1^f$ returns predicted $^{12}$CO/$^{13}$CO values that are in better
agreement with the observed distribution.

With the \citet{watson76} value for $k_1^f$, and keeping all other parameters unchanged
from their chemical modeling results, one is left with a single ingredient that controls
the predicted values of $F_{13}$: the shielding function of CO.
(More accurately, two such ingredients, $\Theta_{12}$ and $\Theta_{13}$.)
Differences between observed $F_{13}$ values and their predictions
are best minimized when both $\Theta_{12}$ and $\Theta_{13}$ are increased by a factor of 2.
It has been suggested that the shielding functions of CO stand in need
for upward revision owing to inconsistencies between isotopic ratio predictions and observations
\citep{sheffer02b,federman03}.
A major contribution to this revision can be provided by an upward revision in the $f$-values
of Rydberg bands of CO, which should enhance both the SPD of CO under low-$N$ conditions,
as well as its self shielding under high-$N$ conditions.
Since the latest determinations of Rydberg $f$-values for all CO isotopologues demonstrate an
upward revision by a factor of $\approx$ 2 \citep{federman01,sheffer03,eidelsberg04},
incorporating these revised $f$-values would presumably help to raise the CO shielding functions
by the same factor.

\subsection{\it $^{12}$CO/$^{13}$CO versus Gas Density}

Table 6 provides values for the gas density, $n_{\rm CN}$, that were derived from our chemical
modeling of $N$(CH) and $N$(CN) along these sight lines.
A chemical network of formation and destruction processes for CN is evaluated for an assumed
value of $I_{\rm UV}$, subject to dust attenuation by the factor exp($-\tau_{\rm UV}$), which
is evaluated at 1000 \AA.
In this modeling, $\tau_{\rm UV}$ has the value 2 $\times$ 3.1 $\times$ $E_{B-V}$, unless extinction
curves and the value of $R_V$ (or $A_V$/$E_{B-V}$) for a specific direction indicate $\sim$ 50\%
change in the prefactor.
This prefactor is about half the value found from extinction curves in order to account for forward
scattering by the dust at far UV wavelengths \citep{federman94}.
The derived values for $n_{\rm CN}$ are found to be proportional to the value of $I_{\rm UV}$,
but they are found to be very insensitive to the input values of $T_{\rm kin}$ of the gas
(a detailed description of the analysis will be given in a future paper).
Although approximate $T_{\rm kin}$ values were used as input (see Table 6), $n_{\rm CN}$ should
change very little by incorporating $T_{1,0}$(H$_2$) from Table 7 instead.

Figure 9 displays the behavior of $^{12}$CO/$^{13}$CO with respect to $n_{\rm CN}$.
The isotopic ratio of CO may show a hint of a mild anti-correlation with $n_{\rm CN}$.
However, inspecting other sight line attributes such as $N$ and $T_{\rm kin}$, we find
no obvious association of values of high gas density with higher CO column densities
(of either isotopologue), nor with lower kinetic temperature of the gas. 
Gas density based on the chemical analysis is expected to be applicable to the CO molecule because
CO and CN are inferred to coexist spatially in diffuse molecular clouds \citep{pan05}.
Another derivation of $I_{\rm UV}$ and $n$ from an analysis of high-$J$ levels of H$_2$ returns
extremely inflated values, as described in $\S$ 4.10.

\subsection{\it $^{12}$CO/$^{13}$CO versus $^{12}$CN/$^{13}$CN}

Other carbon-bearing molecules such as CN also show appreciable variations in fractionation
when probed via (optical) absorption.
As opposed to CO, the CN molecule is dissociated by continuum radiation \citep{lavendy87}
and is not expected to show fractionation to higher values caused by isotopic-selective
photodissociation.
On the other hand, the zero-energy difference of 31 K \citep{bl98} that favors the
heavier $^{13}$CN isotopologue, will work to fractionate CN toward lower 12-to-13 ratios,
especially at low temperatures.
However, when the more abundant CO molecule incorporates most of the $^{13}$C atoms from the
ambient gas, the 12-to-13 ratio of other molecular species, such as CN and H$_2$CO increases,
producing fractionation opposite to that of CO.
Two sight lines offer evidence for this effect happening in CN:
toward $\zeta$ Oph, \citet{rm95} measured $^{12}$CN/$^{13}$CN = 35 $\pm$ 11, whereas
\citet{lambert94} determined $^{12}$CO/$^{13}$CO = 167 $\pm$ 25; as well as toward
HD 154368, where \citet{palazzi90} found the CN 12-to-13 ratio to be 101 $\pm$ 12,
and this paper shows that the CO 12-to-13 ratio is 37 $\pm$ 8.
Thus these two sight lines show CN to have an opposite sense of fractionation to that of CO,
a situation that arises when the more abundant CO controls the availability of $^{13}$C atoms.
As a result of their anti-correlation both CO and CN 12-to-13 values from UV determinations are
seen to vary within a factor-of-2 range about the reference value of $^{12}$C/$^{13}$C = 70.
Simultaneously, $^{12}$CH$^+$/$^{13}$CH$^+$ values are not fractionated: 68 $\pm$ 5 toward
$\zeta$ Oph \citep{crane91} and 58 $\pm$ 8 toward HD 154368 \citep{casassus05}.
Thus the CN and CO isotopic ratios along each sight line
are found on opposite sides of the CH$^+$ (and the ambient carbon) isotopic ratio.
Another sight line suggests a high value of $^{12}$CN/$^{13}$CN, namely 122 $\pm$ 33 toward
HD 21483 \citep{meyer89}, but no information is available concerning the
$^{12}$CO/$^{13}$CO in that direction.
Finally, toward $\zeta$ Per, $^{12}$CO/$^{13}$CO of 108 $\pm$ 5 shows it to be moderately
fractionated, whereas
$^{12}$CN/$^{13}$CN = 77$^{+27}_{-18}$ was given in \citet{kaiser91}.
Given the smaller level of CO fractionation and the relatively large uncertainties in the CN
isotopic ratio,
any final determination will have to be based on data with better S/N.
Interestingly, establishing a relationship of opposite fractionation between CO and CN is tantamount
to showing that both molecules coexist spatially in the ISM, allowing them to interact
with the same pool of $^{13}$C atoms.

\subsection{\it Excitation Temperature of $^{12}$CO and $^{13}$CO}

Each spectrum synthesis with Ismod.f returned fitted values for the excitation temperature
$T_{\rm ex}$ = $T_{J^\prime}$$_{^\prime,0}$ based on the relative populations in the ground state
($J^\prime$$^\prime$ = 0) and in any higher-$J^\prime$$^\prime$ excited level.
The average values of $T_{1,0}$ for both isotopologues agree
with each other very well: $T_{\rm ex}$($^{12}$CO) = 3.7 $\pm$ 0.9 K, and $T_{\rm ex}$($^{13}$CO)
= 3.9 $\pm$ 1.2 K.
The analysis above in $\S$ 2 showed that both isotopologues agree very well in radial velocity,
providing kinematic support for their spatial coexistence.
Having indistinguishable $T_{\rm ex}$ values provides additional strong support that both CO
isotopologues are well mixed and are subject to the same physical conditions.
The smaller samples of \citet{burgh07} and \citet{sonnentrucker07} also indicated
the same $T_{\rm ex}$ for the two CO isotopologues within the stated uncertainties.
The mm-wave absorption survey of CO \citep{ll98b} tabulated $T_{\rm ex}$ values that
provide the averages 6.2 $\pm$ 1.2 and 5.6 $\pm$ 0.7 K for $^{12}$CO and $^{13}$CO,
respectively, also showing them to be identical within the uncertainties.
These mm-wave-derived $T_{\rm ex}$ values are higher than the UV-derived values, suggesting
that denser gas is being observed and that the two samples may be different.

\subsection{\it Lower $T_{1,0}$(H$_2$) along Sight Lines with $^{13}$CO}

The kinetic temperature of the gas is well approximated by $T_{1,0}$(H$_2$), because H$_2$ is
a homonuclear molecule, so that the relative level populations of the ortho and para ground states
are controlled by collisions characterized by $T_{\rm kin}$.
From the 22 sight lines with H$_2$ data in our $^{13}$CO sample, the average
$T_{1,0}$(H$_2$) = 60 $\pm$ 8 K.
This value is lower by 17 and 8 K, respectively, than $T_{1,0}$ averages of 77 $\pm$ 17 K found by
\citet{savage77}, and 68 $\pm$ 15 K from \citet{rachford02}, both of which were H$_2$
surveys unrelated to the presence of CO.
It appears that H$_2$ (and thus, the gas in general) is colder along sight lines with
detectable amounts of $^{13}$CO.
Since all targets in this paper were selected {\it a priori} to show measurable amounts of
$^{13}$CO in their spectra, this selection criterion and the presence of lower kinetic gas
temperatures seem to be related. 
In order to verify that $T_{1,0}$ is lower when $^{13}$CO is present,
we must analyze sight lines from surveys dedicated to CO.

\citet{sonnentrucker07} presented a literature compilation with 38 $T_{1,0}$(H$_2$) values from
CO-bearing sight lines, yielding an average $T_{1,0}$ of 69 $\pm$ 16 K.
However, splitting the sample into sight lines with or without detected amounts of $^{13}$CO,
we find that $T_{1,0}$ = 56 $\pm$ 8 K from 12 directions with $^{13}$CO, whereas $T_{1,0}$ = 76
$\pm$ 15 K from 26 sight lines without detected $^{13}$CO, or $\Delta$$T_{1,0}$ = 20 K.
\citet{burgh07} found an average $T_{1,0}$ of 74 $\pm$ 24 K for 23 CO sight lines.
When we split their sample according to $^{13}$CO detectability, six $^{13}$CO
directions have $T_{1,0}$ = 58 $\pm$ 8 K, while 17 without $^{13}$CO have
$T_{1,0}$ = 80 $\pm$ 25 K, or $\Delta$$T_{1,0}$ = 22 K.
This shows that sight lines with $^{13}$CO from both CO samples have average $T_{1,0}$ values
that are in extremely good agreement with our average of 60 $\pm$ 8 K.
Furthermore, the selection of sight lines with detectable amounts of $^{13}$CO appears to sample
H$_2$ molecules that are colder by $\approx$ 20 K, or equivalently, regions of ambient gas with
$T_{\rm kin}$ that is $\approx$ 20 K lower than along other directions. 

In fact, this clear separation of sight lines into two groups according to their $^{13}$CO
abundance can provide an explanation for the lower $FUSE$ average for $T_{1,0}$ found by
\citet{rachford02} relative to the $Copernicus$ value from \citet{savage77}.
Ten of the 23 sight lines in the \citet{rachford02} sample have detectable amounts of $^{13}$CO;
thus 43\% of that sample is expected to show the presence of colder gas.
The 13 sight lines without $^{13}$CO provide $T_{1,0}$ = 73 $\pm$ 16 K, whereas the
ones with $^{13}$CO have $T_{1,0}$ = 60 $\pm$ 9 K, i.e., a difference of 13 K.
One can observe that the colder value of $T_{1,0}$ is in excellent agreement with the average
$T_{1,0}$ of our $^{13}$CO sight lines, while the warmer value of $T_{1,0}$ agrees nicely with the
average $T_{1,0}$ from \citet{savage77}, which by inspection had 55 lines without
detections of $^{13}$CO, averaging $T_{1,0}$ = 79 $\pm$ 15 K.
This $Copernicus$ sample included only six sight lines currently known to have $^{13}$CO that,
unsurprisingly, yield the low $T_{1,0}$ = 52 $\pm$ 7 K, or 27 K below the value for clouds
without $^{13}$CO.
Since the \citet{rachford02} sample was designed to explore translucent sight lines with higher
$A_V$ and higher $N$(H$_2$), such physical and chemical conditions seem to promote the formation
and survival of detectable amounts of $^{13}$CO.
In fact, the current sample has very similar characteristics to that of \citet{rachford02}: average
value of
$A_V$ is 1.4 $\pm$ 0.5 versus 1.6 $\pm$ 0.7, and average value of log $N$(H$_2$) = 20.7 $\pm$ 0.2
versus 20.8 $\pm$ 0.2.
It is also interesting that \citet{rachford02} remarked that nine $Copernicus$ sight lines with
log $N$(H$_2$) = 20.7 averaged to 55 $\pm$ 8 K, reinforcing the interconnectedness between
higher H$_2$ column density, lower gas kinetic temperature, and the presence of diatomic
molecules with heavy elements such as $^{13}$CO (see below).
The inverse correspondence between the presence of $^{13}$CO and $T_{1,0}$ may be used to predict
detectable abundances of $^{13}$CO along sight lines that have $T_{1,0}$(H$_2$) $\lesssim$ 65 K.

A few of the directions with $^{13}$CO analyzed in this paper were selected from a larger sample
of CO sight lines for which we also determined $T_{1,0}$(H$_2$), based on cloud structures
from newly acquired high-$R$ McD and ESO optical spectra.
For the sake of consistency, the average $T_{\rm kin}$
for the gas along all other sight lines should also be established.
Yet again, 30 directions without $^{13}$CO return
$T_{1,0}$ = 79 $\pm$ 13 K, whereas the four with $^{13}$CO yield $T_{1,0}$ = 59 $\pm$ 10 K, showing
what by now seems to be the ``canonical'' $\Delta$$T_{1,0}$ of $\approx$ 20 K between the two groups
of sight lines as sorted by the presence of $^{13}$CO.

\subsection{\it $^{13}$CO and Other Heavy Diatomic C-bearing Molecules}

We also inspected the literature survey in \citet{sonnentrucker07} according to the presence or
absence of carbon-bearing diatomic molecules other than CO and found that $^{13}$CO is not the
only molecule revealing two groupings of cooler or warmer H$_2$ gas.
Nine sight lines without detected amounts of C$_2$ yielded $T_{1,0}$(H$_2$) = 83 $\pm$ 16 K,
whereas 17 directions with C$_2$ returned the average 57 $\pm$ 9 K, or $\Delta$$T_{1,0}$ = 26 K.
The agreement of $T_{1,0}$ from C$_2$ sight lines with the $^{13}$CO sight lines in $\S$ 4.8 is
outstanding.
As for CN, 10 sight lines without detectable amounts of this molecule average to
$T_{1,0}$(H$_2$) = 79 $\pm$ 13 K, while 28 directions with detections provide 66 $\pm$ 16 K, or
$\Delta$$T_{1,0}$ = 13 K.
These results further enhance the correspondences found between CO, CN, and C$_2$
\citep[e.g.,][]{federman94,pan05,sonnentrucker07}.

The most straightforward connection between $T_{\rm kin}$ of the gas and the abundance of
$^{13}$CO is provided by the exponential dependence of the ICE reaction.
In colder gas the ICE reaction is faster and more efficient in converting
$^{13}$C$^+$ to $^{13}$CO.
Nevertheless, lowering $T_{\rm kin}$ from 80 K along directions without $^{13}$CO to 60 K along
those with $^{13}$CO brings about a decrease of only 15\% in $^{12}$CO/$^{13}$CO.
However, owing to a much more extended sight-line distribution of H$_2$ in comparison with CO,
$T_{1,0}$(H$_2$) may not provide the best indication of $T_{\rm kin}$ for $^{13}$CO.
The more concentrated distribution of heavier molecules has been shown to exist for CN and CO
\citep{pan05}, as well as for C$_2$ \citep{federman94,sonnentrucker07}.
These three molecules contain two heavy elements and are found in denser
and colder clumps of gas with $T_{\rm kin}$ that is better estimated from $T_{\rm ex}$ =
$T_{2,0}$ of rotational populations in C$_2$. 
\citet{sonnentrucker07} tabulated $T_{2,0}$ values for C$_2$ and mentioned that
they are lower than $T_{1,0}$(H$_2$).
Their Table 14 reveals that nine sight lines common to H$_2$ and C$_2$ have average $T_{1,0}$(H$_2$)
of 58 $\pm$ 9 K, whereas the average of $T_{2,0}$(C$_2$) is 28 $\pm$ 11 K, showing that C$_2$ is
colder by 30 K.
If the latter is taken to measure $T_{\rm kin}$ of the gas inside the denser and colder molecular
clumps that also include $^{13}$CO, then one finds a 50 K drop between
environments with and without $^{13}$CO.
This translates into a factor of $\sim$ 2 higher efficiency for the ICE conversion of
$^{13}$C$^+$ into $^{13}$CO, which can nicely account for the those VUV sight lines with
CO that is observed to be fractionated by the same factor.

\subsection{\it Is $^{12}$CO/$^{13}$CO Related to High Excitation Levels of H$_2$?}

The following assumes that the population of higher rotational levels of H$_2$ can be described
by a steady state relation between the formation of rotationally-excited H$_2$ molecules and
UV pumping from the ground state on one hand, and level depopulation by spontaneous radiative
emission on the other.
\citet{lee02} provide a couple of equations that compute the gas density, $n_{\rm tot}$(H),
and the radiation field enhancement, $I_{\rm UV}$, from the observed populations of the
$J^\prime$$^\prime$ = 0, 1, and 4 levels of H$_2$.
These equations are based on the simplified model of \citet{jura74,jura75}, which included a set of
assumptions, such as the lack of dust absorption in H$_2$ Lyman bands ($\sim$ 1000 \AA) and
$n_{\rm tot}$(H) $<$ 10$^4$ cm$^{-3}$, which allows spontaneous emission to proceed without
significant competition from collisional de-excitation.
Table 7 provides hydrogen-related parameters, including Ismod.f modeling values for
$N_{\rm tot}$(H$_2$), $T_{1,0}$, and $T_{4,0}$ (the complete set of results will appear elsewhere).
Although we synthesized the available H$_2$ level populations for 16 sight lines, only 10 of them
were found to have published $N$(H) values, which are needed for $N_{\rm tot}$(H). 
The references for $N$(H) were six sight lines from \citet{ds94}, three sight lines
from \citet{cartledge04}, and a single sight line from \citet{lacour05}.
Also given in Table 7 are values for $I_{\rm UV}$ and $n_{\rm tot}$(H) that were derived using
the equations in \citet{lee02}.

When $^{12}$CO/$^{13}$CO is plotted versus the H$_2$-derived $I_{\rm UV}$
or versus $n_{\rm tot}$(H), there seems to be no discernible correlation.
In fact, since both $n_{\rm tot}$(H) and $I_{\rm UV}$ have much higher derived values than those
determined in the modeling based on CN chemistry (Table 6), the conclusion is that CO is not
associated with regions containing rotationally excited H$_2$.
This confirms that CO is concentrated in denser and colder regions that are much less
extended than the distribution of H$_2$, but are closely related to the conditions inferred
from analysis of CN absorption ($\S$ 4.5).
It is notable that some H$_2$-derived $n_{\rm tot}$(H) values exceed the underlying model
assumption \citep{jura74,jura75} that it be $<$ 10$^4$ cm$^{-3}$, showing that the equations are not
adequate for the analysis of some of these sight lines.
Furthermore, the chemical model explicitly includes dust extinction at 1000 \AA.

A more proper gauge of the enhancement in UV excitation is provided by analyzing
vibrationally-excited absorption lines from H$_2$ \citep{federman95}.
The three sight lines with significant upward CO fractionation are the ones published
before by \citet{federman95} and \citet{federman03}, two of which exhibited detectable
vibrationally-excited H$_2$: $\zeta$ Oph and $\rho$ Oph A.
The predominance of selective photodissociation in the fractionation of CO is thus confirmed
by the presence of vibrationally-excited lines of H$_2$, since searching for the
same lines failed to detect them in other sight lines with unfractionated or downward-fractionated
CO, even those with the highest values of $N$(CO). 

\section{COMPARISON WITH $^{12}$CO/$^{13}$CO FROM OTHER SPECTRAL REGIMES}

As mentioned in $\S$ 4.1, the bottom panel in Fig. 6 shows a histogram of the VUV 12-to-13 results.
Proceeding topward, other histograms show 12-to-13 ratios from optical observations of CH$^+$,
mm-wave absorption observations of CO, mm-wave absorption results from the non-CO molecules
HCO$^+$, HCN, and HNC, and lastly the three mm-wave emission proxies CN, CO, and H$_2$CO.
There seems to be good agreement between sample averages (centering) and the carbon isotopic ratio
of 70 (dashed line) from \citet{wilson99}, which was based on the top two samples.
The mm-wave absorption survey of CO by \citet{ll98b} is responsible for the only
histogram with a different distribution.

Some support for UV-derived CO isotopic ratios that are fractionated upward comes from
the NIR observations of \citet{goto03}, who measured $^{12}$CO/$^{13}$CO of
137 $\pm$ 9 toward LkH$\alpha$~101 and 158 (uncertainty unknown) toward Mon R2 IRS~3, i.e.,
fractionated upward by a factor of $\sim$ 2.
These targets are heavily reddened molecular clouds, illuminated by background IR sources deeply
embedded in star-forming regions, with $N$($^{12}$CO) = 2 $\times$ 10$^{18}$ to 2 $\times$ 10$^{19}$
cm$^{-2}$, or 100 to 1000 times greater than the largest values ($\sim$ 10$^{16}$ cm$^{-2}$) here.
As stated by \citet{goto03}, one would not expect high isotopic values
under conditions of very high extinction that prevent the operation of SPD.
\citet{goto03} note that their fitted $T_{\rm ex}$ values for the cool components of $^{12}$CO
(seen in the transitions with $J$ $\lesssim$ 6) are about twice as high as $T_{\rm ex}$
values of $^{13}$CO, owing to photon trapping in $^{12}$CO transitions.
Since the inferred $N$ values are proportional to $T_{\rm ex}$, the results show isotopic
ratios about twice as high as the ambient value of 70.  

Studies of mm-wave \textit{absorption} toward extragalactic sources have yielded
molecular isotopic ratios appreciably lower than the ambient $^{12}$C/$^{13}$C.
\citet{ll98a} found 59 $\pm$ 2 to be the weighted average of 12-to-13
obtained from absorption profiles of three non-CO carbon-bearing species: HCO$^+$, HCN, and HNC.
The derived average included six absorption components mostly toward a single galactic
direction, whereas one component that yielded H$^{12}$CN/H$^{13}$CN = 170 $\pm$ 51 was
too anomalous to be included in the average.
\citet{ll98a} analyzed five sight lines and found $N$($^{12}$CO)/$N$($^{13}$CO) to range
from 15.3 $\pm$ 2.1 to 54 $\pm$ 13.
Four sight lines from this highly-fractionated sample show at a significance better than 4 $\sigma$
that $^{13}$CO is enhanced by up to a factor of 4.6, which compares favorably with predicted model
values for very cold clouds \citep{vdb88}.
However, as noted by \citet{ll98b}, such mm-wave observations are difficult to interpret
owing to long path lengths and complicated cloud structures.
Their lowest 12-to-13 CO ratios (down to 15) refer to the sight line toward B0355+508,
which presents five kinematic components along a very long path through the molecular layer
of the Galactic disk.
We estimate the upper limit on the path length to be 8.7 kpc from 243 pc/sin(11$\degr$) based on
the 3 $\sigma$ scale height of the molecular component of the ISM \citep{cox05}.
Since this sight line is 150\arcdeg\ away from the direction to the Galactic center,
resulting in an upper limit of 16.6 kpc for $D_{\rm GC}$, its very low $^{12}$CO/$^{13}$CO values
stand in stark contrast to the Galactic gradient of 12-to-13 being proportional to $D_{\rm GC}$
\citep{wr94,wilson99,savage02,milam05}.
We surmise that very cold gas is being probed so that the ICE reaction is even more effective
in producing $^{13}$CO than in more local and warmer gas.
This would be consistent with the findings of \citet{lequeux93} for molecular gas in the outer
Galaxy.

\citet{bensch01}, while reporting the first mm-wave detection of $^{13}$C$^{17}$O emission
in the ISM, also reported C$^{17}$O/$^{13}$C$^{17}$O of 65 $\pm$ 11 toward
the molecular cloud core $\rho$ Oph C, using beam sizes of 25\arcsec\ to 50\arcsec.
This unfractionated value sharply contrasts with the \citet{federman03} VUV value of
$^{12}$CO/$^{13}$CO = 125 $\pm$ 23 in absorption toward the star $\rho$ Oph A.
A way to reconcile this difference in results from the two spectral
regimes is to attribute it to the different volumes being probed by the observing techniques.
The VUV isotopic ratio probes the outer parts of the cloud that absorb the stellar and
interstellar radiation, sampling $N$($^{12}$CO) of $\sim 2 \times 10^{15}$ cm$^{-2}$ that is
sufficient for self shielding by $^{12}$CO, but leaves $^{13}$CO molecules
subjected to high UV flux levels that drive $F_{13}$ up to 1.8.
On the other hand, the mm-wave C$^{17}$O-to-$^{13}$C$^{17}$O ratio tells us that deeper inside the
cloud, where $N$($^{12}$CO), $N$($^{13}$CO), and $N$(C$^{18}$O) are $\approx$ 10$^{19}$, 10$^{17}$,
and 10$^{16}$ cm$^{-2}$, respectively, there is no UV flux and, thus, no CO fractionation.
Since the volume explored by UV absorption lines has a column density $\sim$ 10$^4$ smaller
than that of the volume explored by mm-wave emission, the star $\rho$ Oph A is either
situated in front of the bulk of the CO, or is shining through the outer envelope of the
molecular cloud core.

\section{CONCLUDING REMARKS}

We surveyed the \textit{HST} archive to assemble a sample of 25 diffuse and translucent
(i.e., 0.65 $\leq$ $A_V$ $\leq$ 2.58) sight lines with absorption signatures from $^{13}$CO.
Most of these sight lines are consistent with no fractionation of CO away from the ambient
$^{12}$C/$^{13}$C = 70, while 32\% are found to be significantly fractionated.
Two photochemical processes compete in driving $^{12}$CO/$^{13}$CO away from 70: SPD in
diffuse clouds destroys less of $^{12}$CO thanks to self shielding, whereas the ICE reaction
enhances the abundance of $^{13}$CO in colder clouds.
The former cases are confirmed by the presence of H$_2$ absorption lines that are vibrationally
excited by the enhanced UV radiation field.

Three regimes of $^{12}$CO column density were seen to correspond to varying
contributions from SPD and ICE in affecting CO fractionation.
In the lowest-$N$($^{12}$CO) regime, 12-to-13 values are $\geq$ 70 owing to self shielding
by $^{12}$CO and insignificant competition from ICE.
In the middle regime, where $^{12}$CO/$^{13}$CO revealed the lowest values in the UV, as well as
an overlap with mm-wave absorption results, the presence of C$^+$ allows the ICE reaction to
produce efficiently higher amounts of $^{13}$CO.
Finally, the highest-$N$($^{12}$CO) regime presents the deepest part of the clouds, where
shielding applies to both isotopologues, resulting in $^{12}$CO/$^{13}$CO values close to 70.

The simple model of the fractionation, based on the \citet{lambert94} formula for
$F_{13}$, allowed us to evaluate the basic ingredients that affect fractionation, leading
to a preference of the \citet{watson76} value for the forward reaction between $^{13}$C$^+$
and $^{12}$CO over that published by \citet{sa80} and indicating the need to increase
CO self shielding functions by $\sim$ 2.
However, this simplified model should be followed by more detailed computations
that will include the newly available set of larger (also by $\sim$ 2) $f$-values of CO.

It now appears that the detection of $^{13}$CO can be statistically associated with the presence
of cold gas.
The analysis of $T_{1,0}$(H$_2$) showed that H$_2$ is cooler by about 20 K along sight lines that
possess detectable amounts of $^{13}$CO.
Furthermore, two other carbon-bearing molecules agree with this scenario, since both CN and C$_2$
are also associated with H$_2$ that is colder.
When ICE is the sole fractionating process, $T_{\rm kin}$ of 60 K will establish
an equilibrium value of $^{12}$CO/$^{13}$CO $\approx$ 40, which agrees very well with the
lowest isotopic values in the present VUV sample. 
However, no clear correlation is seen between the 12-to-13 ratio and $T_{\rm kin}$,
presumably because it is masked by the competing route to fractionation, namely SPD, that
affects each sight line to a different extent.

Besides narrow $b$-values that are common to CO and CN absorption lines, the opposite fractionation
relationship that was found between $^{12}$CO/$^{13}$CO and $^{12}$CN/$^{13}$CN also supports the
coexistence of CN and CO.
Together with C$_2$, all three molecules are associated with colder H$_2$ gas.
The inferred coexistence of CO, CN, and C$_2$ in denser and colder portions of a cloud promotes
$T_{2,0}$(C$_2$) instead of $T_{1,0}$(H$_2$) as the better probe of $T_{\rm kin}$ of the
gas in which $^{13}$CO is found.
When $T_{\rm kin}$ $\approx$ 30 K is assumed, the ICE reaction for CO fractionation
will achieve an equilibrium value of $\approx$ 20 in the absence of SPD.
Although we do not see such low values here, similar values have been found for
CO in mm-wave absorption studies \citep{ll98b}.
It remains to be seen whether the mm-wave sample is, indeed, free of UV photodissociation effects,
and whether the VUV CO isotopic ratios happen to hover above $\approx$ 40 owing to competition
from SPD.

Fractionation prevents molecular proxies from perfectly tracking the atomic carbon ratios,
thus direct results concerning the distribution of $^{12}$C and $^{13}$C will eventually have to
come from observations of atomic forms of carbon.
A few steps in this direction were taken by \citet{bb96}, \citet{keene98},
and \citet{tieftrunk01}.
Along two lines of sight only a few arcminutes apart and near the Trapezium cluster at the center
of the Orion Nebula, \citet{bb96} detected 158 $\mu$m emission from both
$^{12}$C$^+$ and $^{13}$C$^+$, whereas \citet{keene98} observed 809 GHz
emission from both $^{12}$C$^0$ and $^{13}$C$^0$.
The two studies yielded the same ratio for $^{12}$C/$^{13}$C: 58 $\pm$ 6 and 58 $\pm$ 12,
respectively.
\citet{keene98} also measured $^{12}$C$^{18}$O/$^{13}$C$^{18}$O = 75 $\pm$ 9.
All three results agree with the solar neighborhood value of 70, considering the quoted
uncertainties.
Still, even $^{12}$C/$^{13}$C may get modified by competition with molecules,
especially CO, for the availability of $^{13}$C, so that a fuller account of the carbon budget
will have to
include all of its isotopes in all atomic and molecular combinations, and in all stages of
ionization, to properly deduce the details of the photochemistry involved.
On the theoretical front, expanding photodissociation and ICE predictions to molecules other
than CO (e.g., CN) would allow a better understanding of isotopic results that are
being assembled for various sight lines.

\acknowledgments
We thank NASA for grant NNG04GD31G and STScI for grant HST-AR-09921.01-A.
Data files for this paper were accessed through the Multiwavelength Archive at STScI.
M. R. acknowledges support by the National Science Foundation under 
Grant No. 0353899 for the Research Experience for Undergraduates 
in the Dept. of Physics and Astronomy at the University of Toledo.
We are grateful to E. Jenkins for sharing STIS data, and to
D. Welty and K. Pan for providing us with cloud structures from optical CH spectra.


\begin{deluxetable}{llllrrcccccc}
\tabletypesize{\scriptsize}
\tablewidth{0pc}
\tablecaption{Stellar Data for $^{12}$CO/$^{13}$CO Sight Lines\tablenotemark{a}}
\tablehead{
\colhead{Star} 
&\colhead{Name} 
&\colhead{Sp.} 
&\colhead{$V$} 
&\colhead{$l$} 
&\colhead{$b$} 
&\colhead{$v_{LSR}\tablenotemark{b}$} 
&\colhead{$E(\bv)$} 
&\colhead{Ref\tablenotemark{c}} 
&\colhead{Distance} 
&\colhead{Ref\tablenotemark{d}} 
&\colhead{$D_{\rm GC}$}\\
\colhead{} 
&\colhead{} 
&\colhead{} 
&\colhead{(mag)} 
&\colhead{(deg)} 
&\colhead{(deg)} 
&\colhead{(km s$^{-1}$)} 
&\colhead{(mag)} 
&\colhead{} 
&\colhead{(pc)} 
&\colhead{} 
&\colhead{(kpc)}}
\startdata
HD  22951 &40 Per      &B0.5 V &4.98 &158.92 &$-$16.70 &$-$6.1 &0.24 &1 & 280 &1 &8.8 \\
HD  23180 &$o$ Per     &B1 III &3.86 &160.36 &$-$17.74 &$-$6.7 &0.30 &2 & 270 &2 &8.8 \\
HD  23478 &            &B3 IV  &6.69 &160.76 &$-$17.42 &$-$6.7 &0.25 &3 & 240 &1 &8.7 \\
HD  24398 &$\zeta$ Per &B1 Iab &2.88 &162.29 &$-$16.69 &$-$7.1 &0.32 &1 & 300 &1 &8.8 \\
HD  24534 &X Per       &O9.5 pe&6.10 &163.08 &$-$17.14 &$-$7.4 &0.59 &4 & 920 &2 &9.4 \\
HD  27778 &62 Tau      &B3 V   &6.33 &172.76 &$-$17.39 &$-$10.2&0.37 &4 & 220 &1 &8.7 \\
HD  96675 &            &B6 IV  &7.6  &296.62 &$-$14.57 &$-$10.7&0.30 &2 & 160 &1 &8.4 \\
HD  99872 &HR 4425     &B3 V   &6.11 &296.69 &$-$10.62 &$-$10.4&0.36 &2 & 230 &1 &8.4 \\
HD 147683 &V760 Sco    &B4 V   &7.05 &344.86 &  +10.09 &7.2    &0.48 &5 & 280 &2 &8.2 \\
HD 147933 &$\rho$ Oph A&B2.5 V &5.02 &353.69 &  +17.69 &10.5   &0.45 &2 & 110 &2 &8.4 \\
HD 148184 &$\chi$ Oph  &B2 Vne &4.42 &357.93 &  +20.68 &11.8   &0.55 &1 & 150 &1 &8.4 \\
HD 148937 &NSV 7808    &O6 f   &6.77 &336.37 &$-$0.22  &3.2    &0.67 &6 &1300 &2 &7.3 \\
HD 149757 &$\zeta$ Oph &O9 V   &2.58 &  6.28 &  +23.59 &14.0   &0.32 &2 & 140 &1 &8.4 \\
HD 154368 &V1074 Sco   &O9 Ia  &6.18 &349.97 &   +3.22 &7.9    &0.78 &4 &1400 &2 &7.1 \\
HD 177989 &            &B2 II  &9.34 & 17.81 &$-$11.88 &12.6   &0.23 &1 &5100 &2 &4.0 \\
HD 192035 &RX Cyg      &B0 IIIn&8.22 & 83.33 &   +7.76 &17.3   &0.28 &1 &2800 &2 &8.6 \\
HD 198781 &HR 7993     &B0.5 V &6.46 & 99.94 &  +12.61 &14.7   &0.35 &2 & 640 &2 &8.6 \\
HD 203374A&            &B0 IVpe&6.69 &100.51 &   +8.62 &14.2   &0.60 &7 & 670 &2 &8.7 \\
HD 203532 &HR 8176     &B3 IV  &6.36 &309.46 &$-$31.74 &$-$8.6 &0.28 &2 & 250 &1 &8.4 \\
HD 206267A&HR 8281     &O6 e   &5.62 & 99.29 &   +3.74 &14.0   &0.53 &4 &1000 &2 &8.7 \\
HD 207198 &HR 8327     &O9 IIe &5.96 &103.14 &   +6.99 &13.5   &0.62 &4 & 990 &2 &8.8 \\
HD 207308 &            &B0.5 V &7.49 &103.11 &   +6.82 &13.4   &0.50 &7 & 810 &2 &8.7 \\
HD 207538 &            &O9 V   &7.30 &101.60 &   +4.67 &13.6   &0.64 &4 & 840 &2 &8.7 \\
HD 208266 &            &B1 V   &8.14 &102.71 &   +4.98 &13.3   &0.52 &7 & 690 &2 &8.7 \\
HD 210839 &$\lambda$ Cep&O6 Iab&5.09 &103.83 &   +2.61 &12.8   &0.57 &2 & 510 &1 &8.6
\enddata
\tablenotetext{a}{We have used the SIMBAD database for non-photometric data.}
\tablenotetext{b}{The correction from heliocentric velocity to the LSR frame.}
\tablenotetext{c}{ (1) \citealt{fruscione94}; (2) \citealt{valencic04}; (3) \citealt{lecoupanec99};
(4) \citealt{jensen05}; (5) \citealt{andersson02}; (6) \citealt{wegner03};
(7) \citealt{simonson68}.}
\tablenotetext{d}{
(1) Parallactic distance from $\geq 4 \sigma$ Hipparcos results;
(2) Spectroscopic parallax based on absolute magnitudes from \citealt{walborn72} and
\citealt{walborn73}.}
\end{deluxetable}


\begin{deluxetable}{lllrrr}
\tabletypesize{\scriptsize}
\tablewidth{0pc}
\tablecaption{\textit{HST} Spectroscopy of Program Stars}
\tablehead{
\colhead{Star}
&\colhead{Data set}
&\colhead{Grating}
&\colhead{Slit}
&\colhead{S/N}
&\colhead{$R$}\\
\colhead{}
&\colhead{}
&\colhead{}
&\colhead{(arcsec)}
&\colhead{}
&\colhead{}}
\startdata
\multicolumn{6}{c}{STIS Data}\\
\hline
HD  22951 &o64805--9  &E140H &0.2X0.05 &105 &121,000 \\ 
HD  23180 &o64801--4  &E140H &0.2X0.05 &100 &119,000 \\
HD  23478 &o6lj01     &E140H &0.1X0.03 &50  &135,000 \\
HD  24398 &o64810--11 &E140H &0.2X0.05 &90  &121,000 \\
HD  24534 &o64812--13 &E140H &0.1X0.03 &105 &140,000 \\
          &o66p01--02 &E140H &0.2X0.09 &110 &101,000 \\
HD  27778 &o59s01     &E140H &0.2X0.09 &40  & 96,000 \\
HD  99872 &o6lj0i     &E140H &0.1X0.03 &35  &139,000 \\
HD 147683 &o6lj06     &E140H &0.2X0.09 &30  &121,000 \\
HD 148937 &o6f301     &E140H &0.2X0.09 &30  &127,000 \\
HD 177989 &o57r03--04 &E140H &0.2X0.09 &75  &104,000 \\
HD 192035 &o6359k     &E140M &0.2X0.2  &35  & 39,000 \\
HD 198781 &o5c049     &E140H &0.2X0.2  &25  & 85,000 \\
HD 203374A &o6lz90    &E140M &0.2X0.2  &45  & 38,000 \\
HD 203532 &o5co1s     &E140H &0.2X0.2  &40  & 80,000 \\
HD 206267A&o5lh09     &E140H &0.1X0.03 &30  &140,000 \\
HD 207198 &o59s06     &E140H &0.2X0.09 &25  &100,000 \\
HD 207308 &o63y02     &E140M &0.2X0.06 &60  & 45,000 \\
HD 207538 &o63y01     &E140M &0.2X0.06 &55  & 47,000 \\
HD 208266 &o63y03     &E140M &0.2X0.06 &60  & 45,000 \\
HD 210839 &o54304     &E140H &0.1X0.03 &45  &160,000 \\
\hline
\multicolumn{6}{c}{GHRS Data}\\
\hline
HD  96675 &z19w01     &G160M &0.25     &30  & 19,000 \\
HD 154368 &z3dw01     &G160M &0.25     &115 & 19,000 \\
          &z0wx01     &G160M &0.25     &20  & 19,000
\enddata
\end{deluxetable}


\begin{deluxetable}{lrcrrcrc}
\tabletypesize{\scriptsize}
\tablewidth{0pc}
\tablecaption{Equivalent Widths and Column Densities for $^{12}$CO}
\tablehead{
\colhead{Star}
&\colhead{$v^{\prime}$}
&\colhead{$\tau <$ 1}
&\colhead{$W_{\lambda}$}
&\colhead{$v^{\prime}$}
&\colhead{$\tau >$ 1}
&\colhead{$W_{\lambda}$}
&\colhead{$N$/10$^{14}$}\\
\colhead{}
&\colhead{}
&\colhead{}
&\colhead{(m\AA)}
&\colhead{}
&\colhead{}
&\colhead{(m\AA)}
&\colhead{(cm$^{-2}$)}}
\startdata
\multicolumn{8}{c}{STIS Data}\\
\hline
HD  22951 & 8 &0.6 & 5.0 $\pm$ 0.2   & 7 &1.3 & 9.3 $\pm$ 0.2 &1.83 $\pm$ 0.06 \\
HD  23180 &\nodata &\nodata &\nodata & 8 &1.5 &15.9 $\pm$ 0.3 &6.78 $\pm$ 0.14 \\
HD  23478 & 9 &0.7 & 9.9 $\pm$ 0.8   & 8 &1.6 &18.9 $\pm$ 0.9 &8.05 $\pm$ 0.52 \\
HD  24398 &\nodata &\nodata &\nodata & 8 &5.4 &24.8 $\pm$ 0.4 &17.9 $\pm$ 0.5  \\
HD  24534 &\nodata &\nodata &\nodata &12 &2.2 &12.1 $\pm$ 0.2 &158. $\pm$ 4.   \\
HD  27778 &13 &0.5 & 5.2 $\pm$ 1.0   &12 &1.2 &12.1 $\pm$ 0.9 &123. $\pm$ 17.  \\
HD  99872 & 9 &0.7 & 5.5 $\pm$ 0.6   & 8 &1.6 &10.2 $\pm$ 0.8 &4.54 $\pm$ 0.43 \\ 
HD 147683 &13 &0.4 & 3.8 $\pm$ 0.8   &12 &1.1 & 8.8 $\pm$ 0.9 &80.3 $\pm$ 12.6 \\
HD 148937 & 8 &0.8 &10.2 $\pm$ 2.5   & 7 &1.7 &18.6 $\pm$ 2.9 &3.81 $\pm$ 0.76 \\
HD 177989 &10 &0.4 & 2.4 $\pm$ 0.8   & 9 &1.0 & 5.2 $\pm$ 0.6 &4.4  $\pm$ 1.0  \\
HD 192035 &10 &0.7 & 7.1 $\pm$ 1.6   & 9 &1.6 &14.4 $\pm$ 1.9 &13.9 $\pm$ 2.5  \\
HD 198781 &11 &0.7 & 3.6 $\pm$ 0.8   &10 &1.6 & 7.1 $\pm$ 0.5 &16.6 $\pm$ 2.5  \\
HD 203374A&11 &0.7 & 5.7 $\pm$ 1.2   &10 &1.6 &11.5 $\pm$ 1.4 &25.5 $\pm$ 4.3  \\
HD 203532 &12 &0.5 & 5.0 $\pm$ 0.7   &11 &1.1 & 9.1 $\pm$ 0.8 &45.6 $\pm$ 5.3  \\
HD 206267A&\nodata &\nodata &\nodata &12 &1.6 &12.3 $\pm$ 2.1 &134. $\pm$ 23.  \\
HD 207198 &11 &0.6 & 7.2 $\pm$ 0.8   &10 &1.4 &15.1 $\pm$ 0.5 &31.6 $\pm$ 2.1  \\
HD 207308 &12 &0.9 & 8.6 $\pm$ 1.5   &11 &1.9 &15.2 $\pm$ 1.1 &83.2 $\pm$ 10.4 \\
HD 207538 &11 &0.5 & 5.5 $\pm$ 0.7   &10 &1.1 &11.6 $\pm$ 0.9 &23.4 $\pm$ 2.4  \\
HD 208266 &13 &0.6 & 4.8 $\pm$ 1.6   &12 &1.7 &10.7 $\pm$ 0.9 &116. $\pm$ 24.  \\
HD 210839 &11 &0.9 & 5.9 $\pm$ 0.4   &10 &2.2 &11.5 $\pm$ 0.3 &27.5 $\pm$ 1.3  \\
\hline
\multicolumn{8}{c}{GHRS Data}\\
\hline
HD  96675 &\nodata &\nodata &\nodata & 8 &4.5 &31.0 $\pm$ 2.9 &20.2 $\pm$ 3.2 \\
HD 154368 &10 &0.9 &11.9 $\pm$ 3.1   & 9 &2.2 &23.5 $\pm$ 3.5 &26.7 $\pm$ 5.5
\enddata
\end{deluxetable}


\begin{deluxetable}{lccrccrc}
\tabletypesize{\scriptsize}
\tablewidth{0pc}
\tablecaption{Equivalent Widths and Column Densities for $^{13}$CO}
\tablehead{
\colhead{Star}
&\colhead{$v^{\prime}$}
&\colhead{$\tau <$ 1}
&\colhead{$W_{\lambda}$}
&\colhead{$v^{\prime}$}
&\colhead{$\tau >$ 1}
&\colhead{$W_{\lambda}$}
&\colhead{$N$/10$^{12}$}\\
\colhead{}
&\colhead{}
&\colhead{}
&\colhead{(m\AA)}
&\colhead{}
&\colhead{}
&\colhead{(m\AA)}
&\colhead{(cm$^{-2}$)}}
\startdata
\multicolumn{8}{c}{STIS Data}\\
\hline
HD  22951 & 2 &0.2 & 1.7 $\pm$ 0.2 &\nodata &\nodata &\nodata &2.31 $\pm$ 0.33 \\
HD  23180 & 2 &0.4 & 6.2 $\pm$ 0.4 &\nodata &\nodata &\nodata &9.32 $\pm$ 0.55 \\
HD  23478 & 4 &0.8 & 4.9 $\pm$ 0.9 & 3 &1.2 & 6.9 $\pm$ 0.9   &12.0 $\pm$ 1.9 \\
HD  24398 & 3 &0.9 & 8.2 $\pm$ 0.3 & 2 &1.0 & 9.5 $\pm$ 0.3   &16.5 $\pm$ 0.6 \\
HD  24534 & 9 &0.5 & 2.4 $\pm$ 0.2 & 8 &1.0 & 4.7 $\pm$ 0.1   &186. $\pm$ 9. \\
HD  27778 & 8 &0.5 & 4.8 $\pm$ 0.4 & 7 &1.1 & 9.1 $\pm$ 0.5   &183. $\pm$ 12. \\
HD  99872 & 2 &0.4 & 3.9 $\pm$ 0.9 &\nodata &\nodata &\nodata &5.6  $\pm$ 1.3 \\ 
HD 147683 & 8 &0.3 & 3.2 $\pm$ 0.4 & 7 &0.6 & 6.3 $\pm$ 0.5   &113. $\pm$ 11. \\
HD 148937 & 2 &0.3 & 4.4 $\pm$ 1.5 &\nodata &\nodata &\nodata &6.6  $\pm$ 2.3 \\
HD 177989 & 3 &0.7 & 4.4 $\pm$ 0.5 & 2 &0.8 & 5.2 $\pm$ 0.5   &8.35 $\pm$ 0.88 \\
HD 192035 & 3 &0.9 &14.0 $\pm$ 1.9 & 2 &1.1 &16.4 $\pm$ 1.7   &27.5 $\pm$ 3. \\
HD 198781 & 7 &0.3 & 1.7 $\pm$ 0.6 &\nodata &\nodata &\nodata &27.7 $\pm$ 10.4\\
HD 203374A& 5 &0.6 & 6.9 $\pm$ 1.6 & 4 &1.0 &11.1 $\pm$ 1.5   &33.3 $\pm$ 6.1 \\
HD 203532 & 8 &0.4 & 2.8 $\pm$ 0.4 & 7 &0.9 & 5.3 $\pm$ 0.5   &111. $\pm$ 12. \\
HD 206267A& 8 &0.8 & 7.9 $\pm$ 1.3 & 7 &1.7 &14.0 $\pm$ 1.8   &322. $\pm$ 47. \\
HD 207198 & 7 &0.3 & 4.0 $\pm$ 0.6 &\nodata &\nodata &\nodata &66.1 $\pm$ 9.7 \\
HD 207308 & 7 &0.7 & 7.0 $\pm$ 0.9 & 6 &1.4 &12.5 $\pm$ 0.9   &118. $\pm$ 12. \\
HD 207538 & 5 &0.7 & 9.9 $\pm$ 0.8 & 4 &1.3 &15.8 $\pm$ 0.7   &45.6 $\pm$ 2.7 \\
HD 208266 & 7 &0.8 & 7.8 $\pm$ 1.0 & 6 &1.6 &13.7 $\pm$ 0.8   &138. $\pm$ 12. \\
HD 210839 & 7 &0.3 & 1.7 $\pm$ 0.3 &\nodata &\nodata &\nodata &35.4 $\pm$ 6.2\\
\hline
\multicolumn{8}{c}{GHRS Data}\\
\hline
HD  96675 &\nodata &\nodata &\nodata & 3 &1.3 &11.7 $\pm$ 1.8 &24.9 $\pm$ 3.9 \\
HD 154368 & 5 &0.7 &12.9 $\pm$ 2.7 & 4 &1.2 &20.5 $\pm$ 1.0   &71.3 $\pm$ 9.2
\enddata
\end{deluxetable}


\begin{deluxetable}{lclllll}
\tabletypesize{\scriptsize}
\tablewidth{0pc}
\tablecaption{Comparison of Column Densities and Ratios with Published Results}
\tablehead{
\colhead{Star}
&\colhead{CO}
&\multicolumn{5}{l}{log $N$ (cm$^{-2}$) and $^{12}$CO/$^{13}$CO values:}\\
\colhead{}
&\colhead{}
&\multicolumn{1}{l}{This Paper:}
&\multicolumn{4}{l}{Previous References:}
}
\startdata
HD  22951 &12 &14.26 $\pm$ 0.01 &(1) 14.22            &&& \\
HD  23180 &12 &14.83 $\pm$ 0.01 &(1) 14.93            &&& \\
HD  24534 &12 &16.20 $\pm$ 0.01 &(2) 16.00 $\pm$ 0.08 &(3) 16.15 $\pm$ 0.06 &(4) 16.01 $\pm$ 0.09 &(5) 16.13 $\pm$ 0.20 \\
HD  27778 &12 &16.09 $\pm$ 0.06 &(4) 16.08 $\pm$ 0.03 &(5) 16.05 $\pm$ 0.13 && \\
HD 177989 &12 &14.64 $\pm$ 0.09 &(5) 14.62 $\pm$ 0.17 &&& \\
HD 203374A&12 &15.41 $\pm$ 0.07 &(6) 15.38 $\pm$ 0.04 &(7) 15.41 $\pm$ 0.02 && \\
HD 203532 &12 &15.66 $\pm$ 0.05 &(5) 15.70 $\pm$ 0.17 &&& \\
HD 206267A&12 &16.13 $\pm$ 0.12 &(4) 16.04 $\pm$ 0.04 &(5) 16.11 $\pm$ 0.17 &(7) 16.00 $\pm$ 0.03 & \\
HD 207198 &12 &15.50 $\pm$ 0.03 &(4) 15.52 $\pm$ 0.04 &(7) 15.42 $\pm$ 0.02 && \\
HD 207308 &12 &15.92 $\pm$ 0.05 &(7) 15.93 $\pm$ 0.05 &&& \\
HD 207538 &12 &15.37 $\pm$ 0.04 &(7) 15.40 $\pm$ 0.04 &&& \\
HD 208266 &12 &16.06 $\pm$ 0.08 &(7) 16.05 $\pm$ 0.04 &&& \\
HD 210839 &12 &15.44 $\pm$ 0.02 &(4) 15.46 $\pm$ 0.06 &(5) 15.41 $\pm$ 0.04 &(7) 15.39 $\pm$ 0.03 & \\
\hline
HD 24534  &13 &14.27 $\pm$ 0.02 &(2) 14.43            &(3) 14.29 $\pm$ 0.02 &(4) 14.23 $\pm$ 0.03 &(5) 14.30 $\pm$ 0.12 \\
HD 27778  &13 &14.26 $\pm$ 0.03 &(4) 14.19 $\pm$ 0.10 &(5) 14.28 $\pm$ 0.08 && \\
HD 177989 &13 &12.92 $\pm$ 0.04 &(5) 12.82 $\pm$ 0.08 &&& \\
HD 203532 &13 &14.05 $\pm$ 0.08 &(5) 13.97 $\pm$ 0.20 &&& \\
HD 206267A&13 &14.51 $\pm$ 0.06 &(4) 14.39 $\pm$ 0.03 &(5) 14.42 $\pm$ 0.08 && \\
HD 207198 &13 &13.82 $\pm$ 0.06 &(4) 13.75 $\pm$ 0.10 &&& \\
HD 210839 &13 &13.55 $\pm$ 0.07 &(4) 13.59 $\pm$ 0.08 &(5) 13.70 $\pm$ 0.10 && \\
\hline
HD 24534  &12/13 &85 $\pm$  5 &(2) 37          &(3) 73 $\pm$ 12 &(4) 60 $\pm$ 13 &(5) 68 $\pm$ 31 \\
HD 27778  &12/13 &67 $\pm$ 10 &(4) 79 $\pm$ 12 &(5) 59 $\pm$ 14 && \\
HD 177989 &12/13 &53 $\pm$ 13 &(5) 63 $\pm$ 25 &&& \\
HD 203532 &12/13 &41 $\pm$  7 &(5) 54 $\pm$ 21 &&& \\
HD 206267A&12/13 &42 $\pm$  9 &(4) 46 $\pm$  6 &(5) 49 $\pm$ 15 && \\
HD 207198 &12/13 &48 $\pm$  8 &(4) 59 $\pm$ 14 &&& \\
HD 210839 &12/13 &78 $\pm$ 14 &(4) 74 $\pm$ 17 &(5) 51 $\pm$  9 &&
\enddata
\tablerefs{(1) \citealt{wannier99}; (2) \citealt{kaczmarczyk00}; (3) \citealt{sheffer02a};
(4) \citealt{sonnentrucker07}; (5) \citealt{burgh07}; (6) \citealt{sheffer03}; (7) \citealt{pan05}.}
\end{deluxetable}


\begin{deluxetable}{lcccccrll}
\tabletypesize{\scriptsize}
\tablewidth{0pc}
\tablecaption{CO Isotopic Ratios and Fractionation Parameters}
\tablehead{
\colhead{Star}
&\colhead{$^{12}$CO/$^{13}$CO}
&\colhead{$F_{13}$}
&\colhead{$T_{\rm kin}$}
&\colhead{$I_{\rm UV}$}
&\colhead{$\tau_{\rm UV}$}
&\colhead{$n_{\rm CN}$\tablenotemark{a}}
&\colhead{$\Theta_{12}$}
&\colhead{$\Theta_{13}$}\\
\colhead{}
&\colhead{}
&\colhead{}
&\colhead{(K)}
&\colhead{}
&\colhead{}
&\colhead{(cm$^{-3}$)}
&\colhead{}
&\colhead{}}
\startdata
\multicolumn{9}{c}{STIS Data}\\
\hline
HD  22951 & 79 $\pm$ 12 &1.13 &40 &1.0 &1.49 &225 &0.473 &0.563 \\
HD  23180 & 73 $\pm$  5 &1.04 &40 &1.0 &1.86 &625 &0.298 &0.450 \\
HD  23478 & 67 $\pm$ 13 &0.96 &50 &1.0 &1.67 &525 &0.275 &0.435 \\
HD  24398 &108 $\pm$  5 &1.54 &30 &1.0 &2.05 &700 &0.189 &0.362 \\
HD  24534 & 85 $\pm$  5 &1.21 &20 &1.0 &3.84 &650 &0.0372&0.154 \\
HD  27778 & 67 $\pm$ 10 &0.96 &50 &0.5 &2.29 &$\sim$ 900 &0.0465&0.176 \\
HD  99872 & 81 $\pm$ 21 &1.16 &65 &1.0 &1.90 &$\leq$200 &0.342 &0.494 \\
HD 147683 & 71 $\pm$ 13 &1.01 &\nodata &\nodata &\nodata &\nodata &\nodata &\nodata \\
HD 148937 & 58 $\pm$ 23 &0.83 &\nodata &\nodata &\nodata &\nodata &\nodata &\nodata \\
HD 177989 & 53 $\pm$ 13 &0.76 &\nodata &\nodata &\nodata &\nodata &\nodata &\nodata \\
HD 192035 & 51 $\pm$ 10 &0.73 &65 &1.0 &2.05 &$\sim$ 1450 &0.221 &0.386 \\
HD 198781 & 60 $\pm$ 23 &0.86 &65 &1.0 &3.26 &750 &0.217 &0.396 \\
HD 203374A& 77 $\pm$ 19 &1.10 &50 &1.0 &3.69 &80  &0.160 &0.333 \\
HD 203532 & 41 $\pm$  7 &0.59 &\nodata &\nodata &\nodata &\nodata &\nodata &\nodata \\
HD 206267A& 42 $\pm$  9 &0.60 &50 &1.0 &3.13 &1000 &0.0372&0.154 \\
HD 207198 & 48 $\pm$  8 &0.69 &50 &1.0 &3.63 &130 &0.128 &0.290 \\
HD 207308 & 71 $\pm$ 11 &1.01 &50 &1.0 &3.00 &550 &0.0617&0.198 \\
HD 207538 & 51 $\pm$  6 &0.73 &50 &1.0 &3.92 &100 &0.144 &0.298 \\
HD 208266 & 84 $\pm$ 19 &1.20 &50 &1.0 &3.10 &300 &0.0440&0.167 \\
HD 210839 & 78 $\pm$ 14 &1.11 &50 &1.0 &2.36 &375 &0.151 &0.316 \\
\hline
\multicolumn{9}{c}{GHRS Data}\\
\hline
HD  96675 & 81 $\pm$ 18 &1.16 &50 &1.0 &1.58 &1425 &0.170 &0.324 \\
HD 147933 &125 $\pm$ 36 &1.79 &50 &1.0 &2.04 &350 &0.200 &0.383 \\
HD 148184 &117 $\pm$ 55 &1.67 &60 &1.0 &2.30 &$\sim$ 300 &0.350 &0.480 \\
HD 149757 &167 $\pm$ 25 &2.39 &60 &1.0 &1.98 &325 &0.160 &0.333 \\
HD 154368 & 37 $\pm$ 8  &0.53 &50 &1.0 &4.77 &750 &0.122 &0.253
\enddata
\tablenotetext{a}{Average gas density from chemical modeling, weighted by observed CN component fractions.}
\end{deluxetable}


\begin{deluxetable}{lcccrr}
\tabletypesize{\scriptsize}
\tablewidth{0pc}
\tablecaption{H$_2$-Related Observables}
\tablehead{
\colhead{Star}
&\colhead{log $N$(H$_2$)}
&\colhead{$T_{\rm 0,1}$(H$_2$)}
&\colhead{$T_{\rm 0,4}$(H$_2$)}
&\colhead{$I_{\rm UV}$\tablenotemark{a}}
&\colhead{$n_{\rm H_2}$\tablenotemark{a}}\\
\colhead{}
&\colhead{(cm$^{-2}$)}
&\colhead{(K)}
&\colhead{(K)}
&\colhead{}
&\colhead{(cm$^{-3}$)}}
\startdata
HD  22951  &20.46 &63 &\nodata &\nodata &\nodata \\
HD  23180  &20.61 &48 &\nodata &\nodata &\nodata \\
HD  23478  &20.57 &55 &171  &\nodata &\nodata \\
HD  24398  &20.67 &57 &$\geq$ 117 &\nodata &\nodata \\
HD  24534  &20.94 &54 &152  &95 &1900 \\
HD  27778  &20.79 &51 &152  &80 &1230 \\
HD  96675  &20.87 &55 &164  &\nodata &\nodata \\
HD  99872  &20.51 &66 &179  &\nodata &\nodata \\
HD 147683  &20.74 &58 &185  &\nodata &\nodata \\
HD 147933  &20.57 &46 &\nodata &\nodata &\nodata \\
HD 148184  &20.63 &46 &\nodata &\nodata &\nodata \\
HD 148937  &20.71 &69 &228  &\nodata &\nodata \\
HD 149757  &20.65 &54 &$\geq$ 129 &\nodata &\nodata \\
HD 154368  &21.16 &46 &140  &\nodata &\nodata \\
HD 177989  &20.15 &49 &198  &\nodata &\nodata \\
HD 192035  &20.68 &68 &205  &1200 &18400 \\
HD 198781  &20.56 &65 &191  &570 &10900 \\
HD 203374A &20.70 &76 &135  &17 &260 \\
HD 203532  &20.70 &47 &169  &220 &2600 \\
HD 206267A &20.86 &58 &156  &110 &1350 \\
HD 207198  &20.83 &60 &162  &168 &1900 \\
HD 207308  &20.86 &57 &162  &\nodata &\nodata \\
HD 207538  &20.91 &66 &145  &53 &610 \\
HD 208266  &[20.94]\tablenotemark{b} &\nodata &\nodata  &\nodata &\nodata \\
HD 210839  &20.84 &69 &218  &2300 &33000
\enddata
\tablenotetext{a}{Radiation field strength and gas density as implied by UV excitation of $N_{\rm J=4}$(H$_2$).}
\tablenotetext{b}{$N$(H$_2$) for HD 208266 is predicted from $N$(CH) and $N$(CO), see Table 2 in
\citealt{pan05}.}
\end{deluxetable}


\begin{figure}
\epsscale{1.0}
\plotone{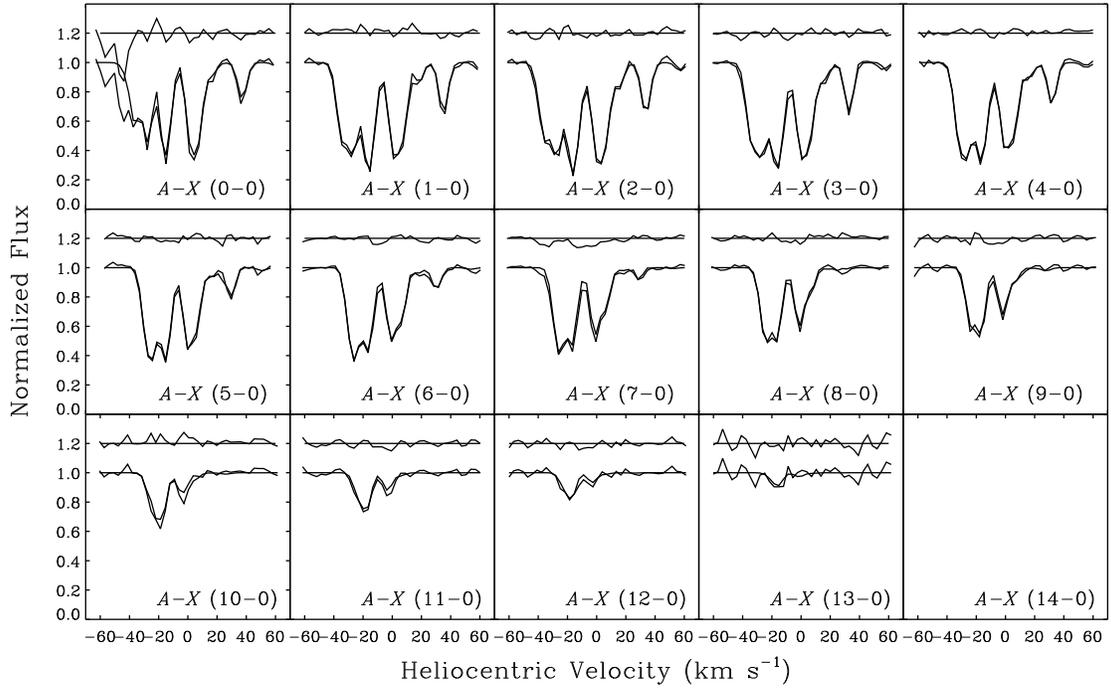}
\caption{Example of a complete simultaneous fit to all 14 detectable $A$--$X$ bands
of $^{12}$CO redward of Ly$\alpha$ along the sight line of HD 208266.
This complete detection is possible thanks to the large value of $N$($^{12}$CO)
and to the extensive ($>$315 \AA) coverage of the STIS E140M setup.
The residual absorption on the blue side of the (0--0) band is
from $^{13}$CO, of which eight $A$--$X$ bands are modeled in Fig. 2.
The noise is appreciably larger at the (13--0) band, which is $\sim$ 14 \AA\ redward of the
Ly$\alpha$ line center.
The undetected (14--0) band is weaker than (13--0) by a factor of 2,
and is completely lost in the noise inside the Ly$\alpha$ core.
The quality of the fit is indicated by the variation of residuals about the 1.2 level in
this figure and in Fig. 2.}
\end{figure}


\begin{figure}
\epsscale{1.0}
\plotone{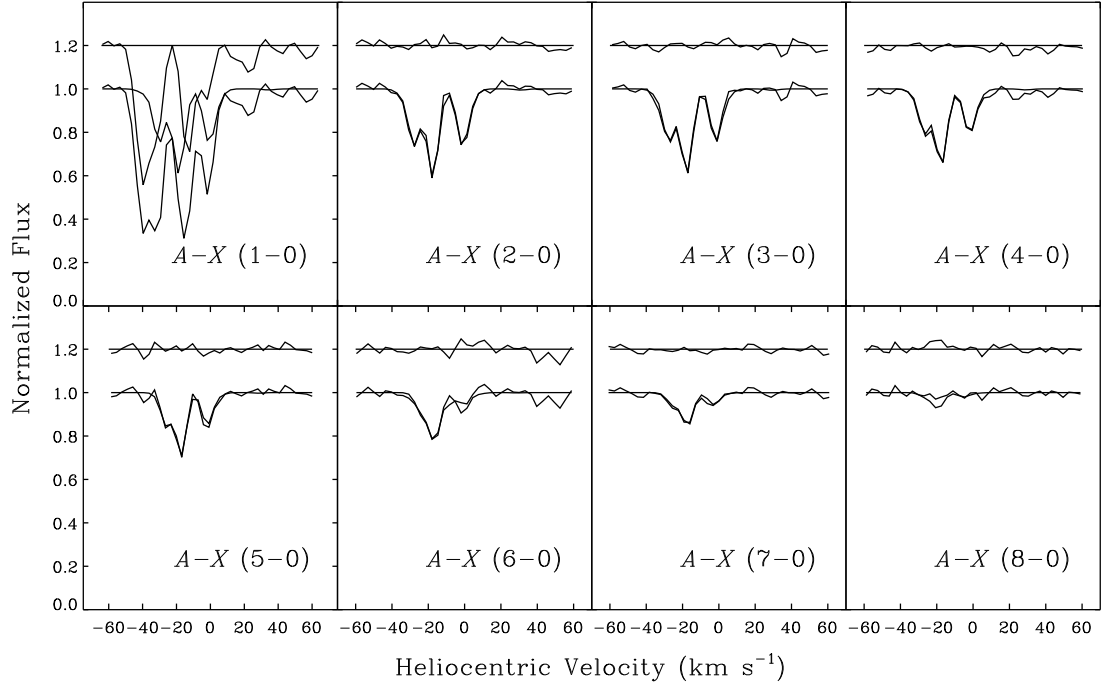}
\caption{Simultaneous fit to 8 $A$--$X$ bands of $^{13}$CO for HD 208266.
The prominent absorption feature blended with the (1--0) band is the $d$--$X$ (5--0)
band of $^{12}$CO, which is the strongest intersystem transition in CO.
For this reason, the (1--0) band was not incorporated into the fit, yet it shows good agreement
with the global model.}
\end{figure}


\begin{figure}
\epsscale{0.7}
\plotone{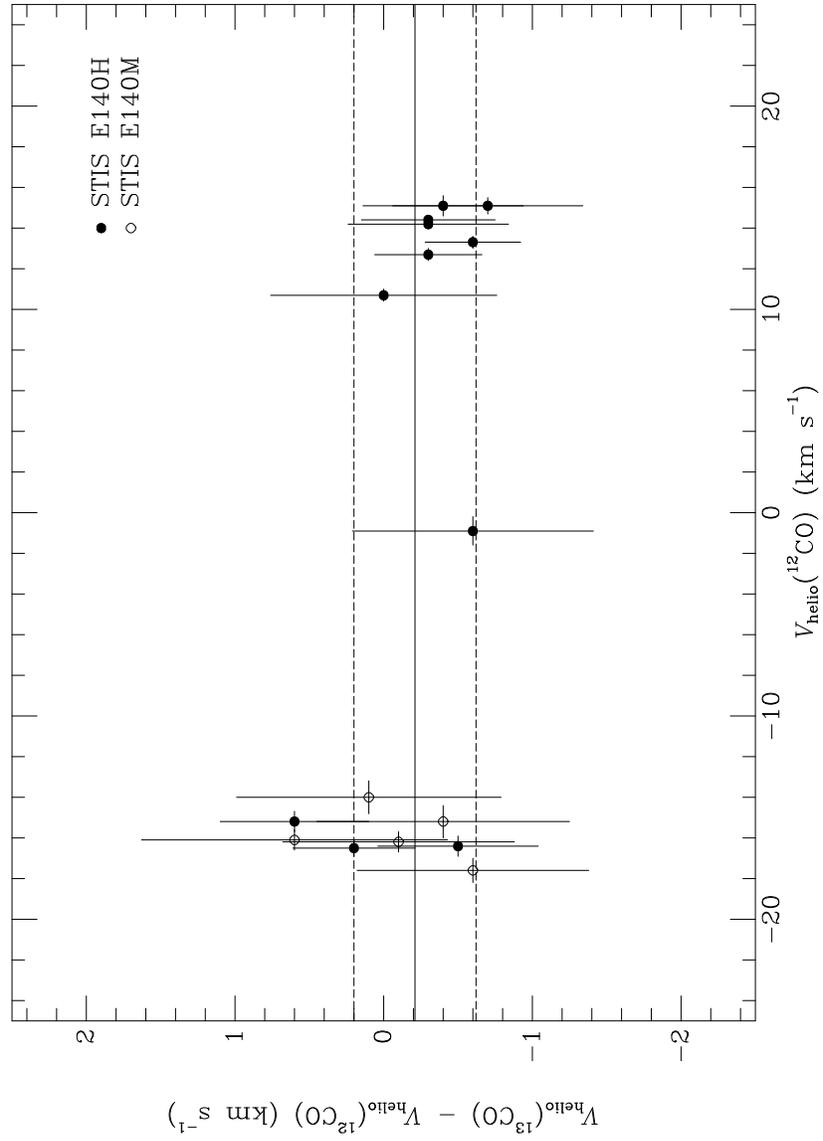}
\caption{Difference in radial velocity between $^{13}$CO and $^{12}$CO for STIS sight lines
shows the good agreement between the two species, and hence supports a physical association between
them.
No difference is seen between results from the E140H and E140M gratings.
GHRS spectra have larger wavelength calibration uncertainties and their results are not plotted.}
\end{figure}


\begin{figure}
\epsscale{1.0}
\plotone{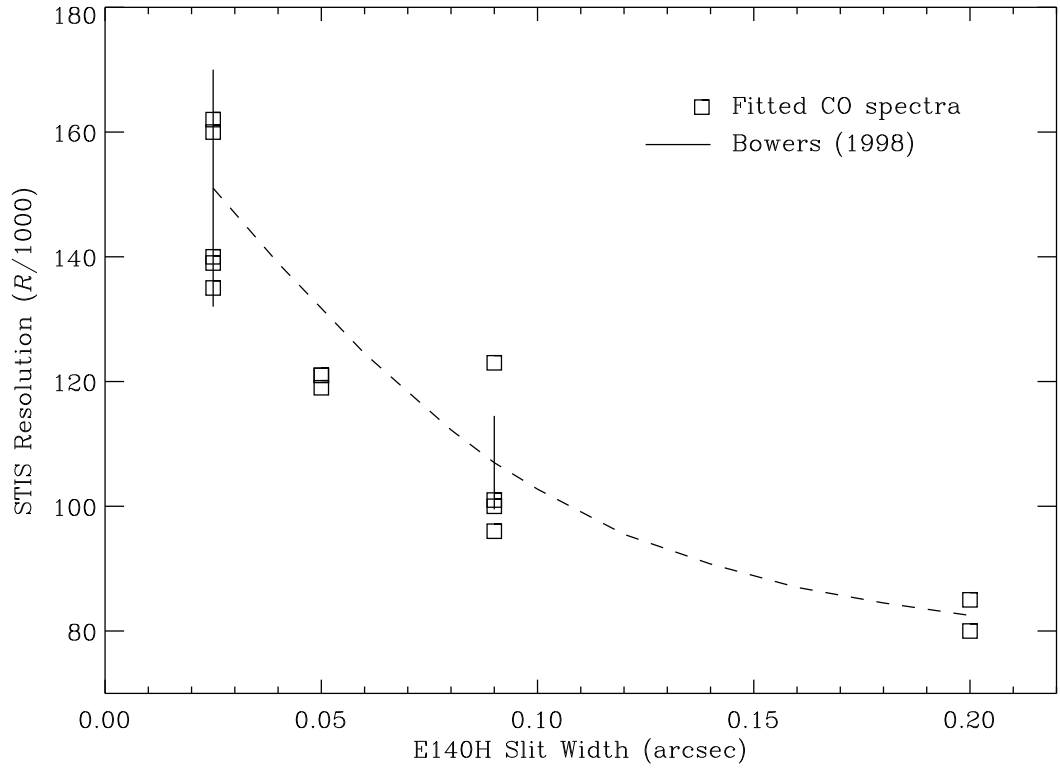}
\caption{Resolving power ($R$) from our fits to STIS spectra is plotted against the slit width
used with the E140H grating.
Note that the y-scale corresponds to the value of $R$/1000.
A freely drawn dashed line shows the non-linear trend of $R$ as a function of slit width.
Two ``error bars'' show the range of results from \citet{bowers98}, which agree nicely with our
determinations.}
\end{figure}


\begin{figure}
\epsscale{1.0}
\plotone{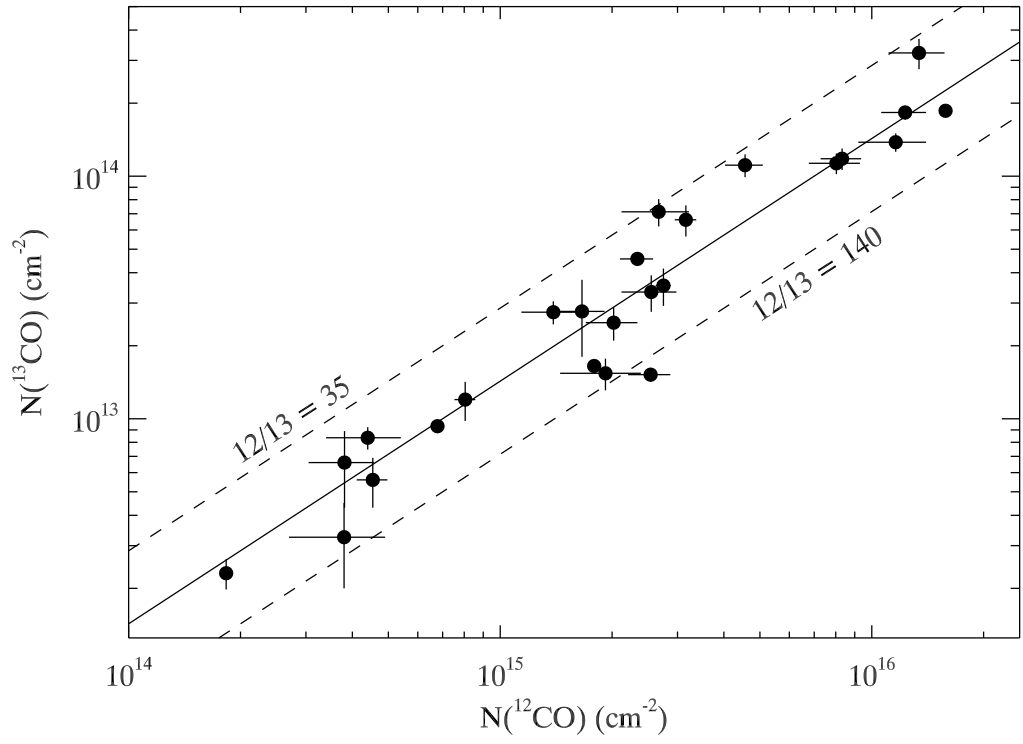}
\caption{Measured column densities for both $^{12}$CO and $^{13}$CO.
For comparison, $^{12}$C/$^{13}$C = 70 is depicted by the solid line.
Except for one sight line, 96\% of the sample have fractionation values within a factor of 2 of 70,
as delineated by the dashed lines corresponding to $^{12}$CO/$^{13}$CO
= 35 and 140, or $F_{13}$ = 0.5 and 2.0.}
\end{figure}


\begin{figure}
\epsscale{1.0}
\plotone{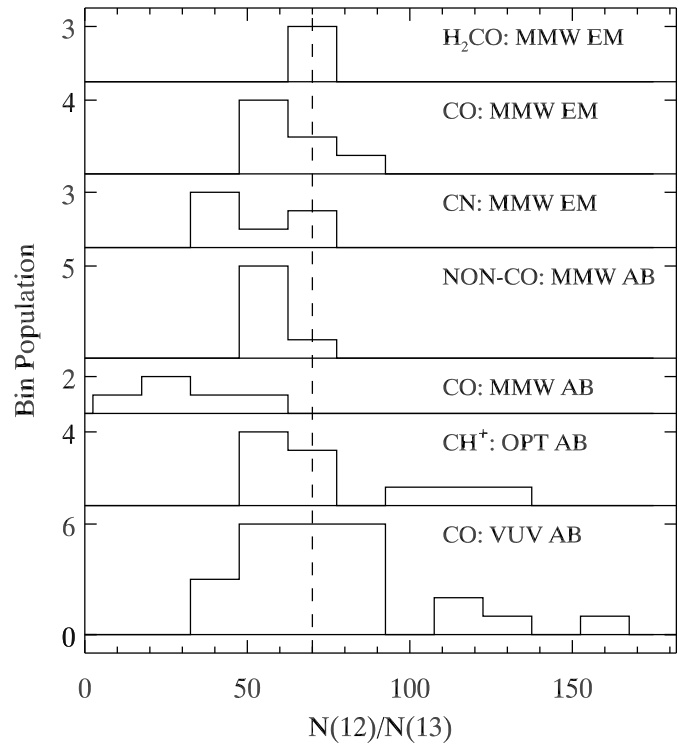}
\caption{Histograms of 12-to-13 carbon ratios from absorption (ab) or emission (em)
surveys conducted in various spectral regimes.
A reference bin has been centered at 70 $\pm$ 7, with all other bins having the same 2 $\sigma$
width.}
\end{figure}


\begin{figure}
\epsscale{1.0}
\plotone{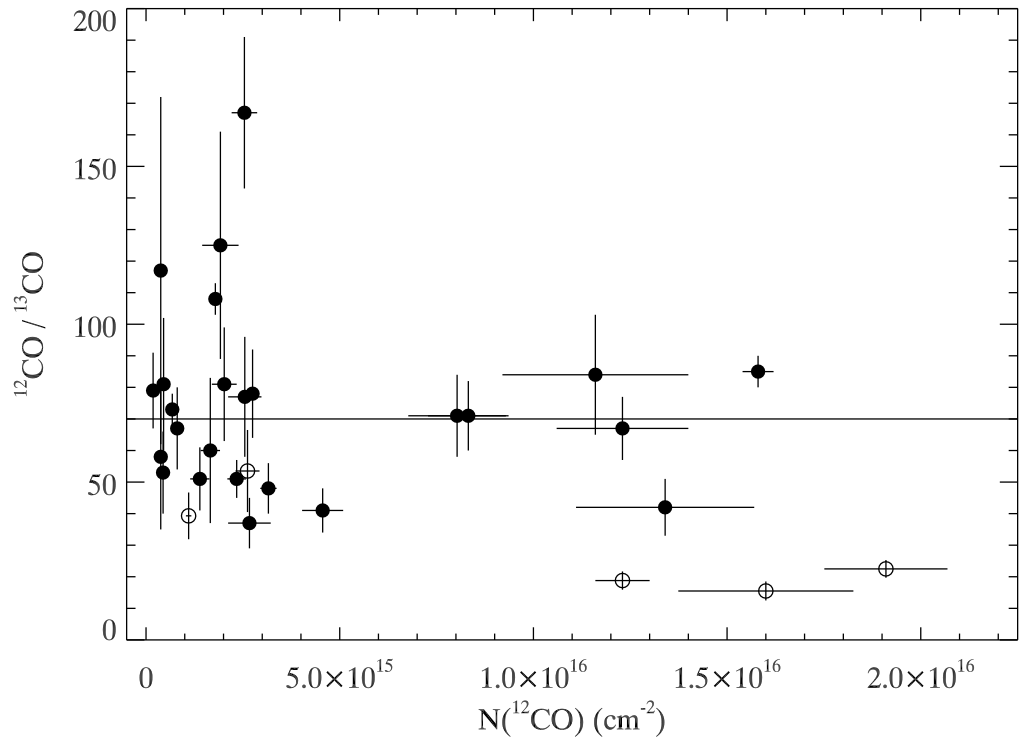}
\caption{CO isotopic ratio versus column density of $^{12}$CO.
Filled symbols denote UV determinations, while empty symbols stand for the mm-wave absorption
measurements of \citet{ll98b}.
The CO along the latter sight lines has been summed for all velocity components for the
sake of consistency with the VUV results.
A solid line shows the adopted isotopic carbon ratio of 70.}
\end{figure}


\begin{figure}
\epsscale{1.0}
\plotone{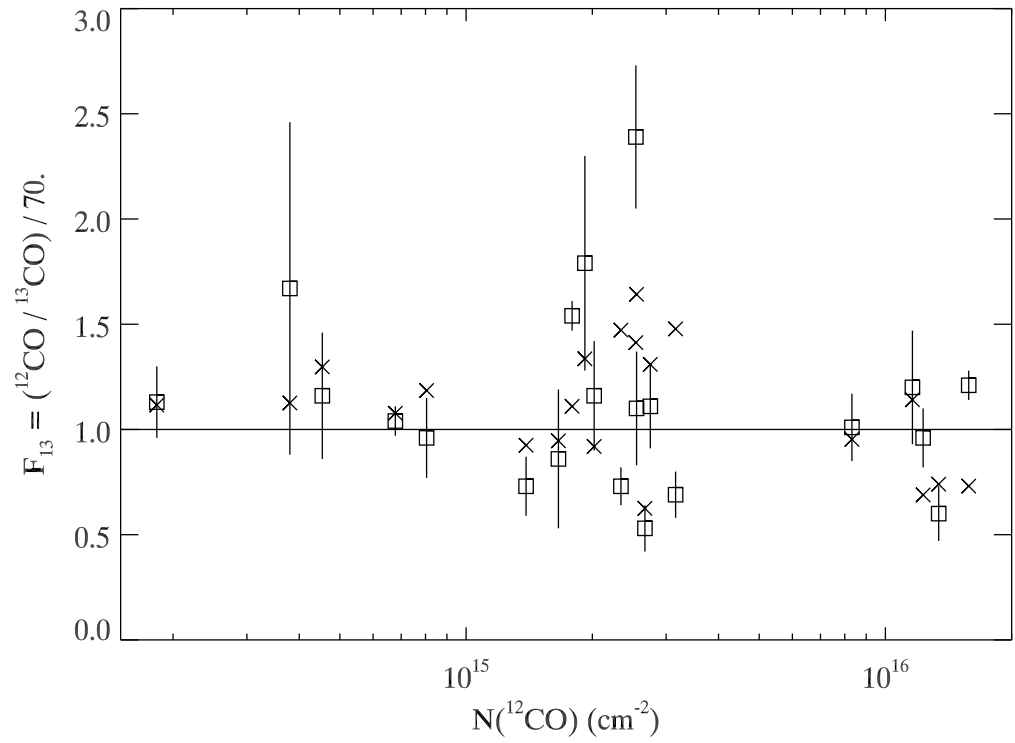}
\caption{Observed CO isotopic ratios, normalized by the carbon isotopic ratio, are plotted versus
log $N$($^{12}$CO) as empty squares.
A comparison is made with fractionation predictions (denoted by X symbols) based on the
\citet{lambert94} formula and on tabulated shielding values from \citet{vdb88}.}
\end{figure}


\begin{figure}
\epsscale{0.7}
\plotone{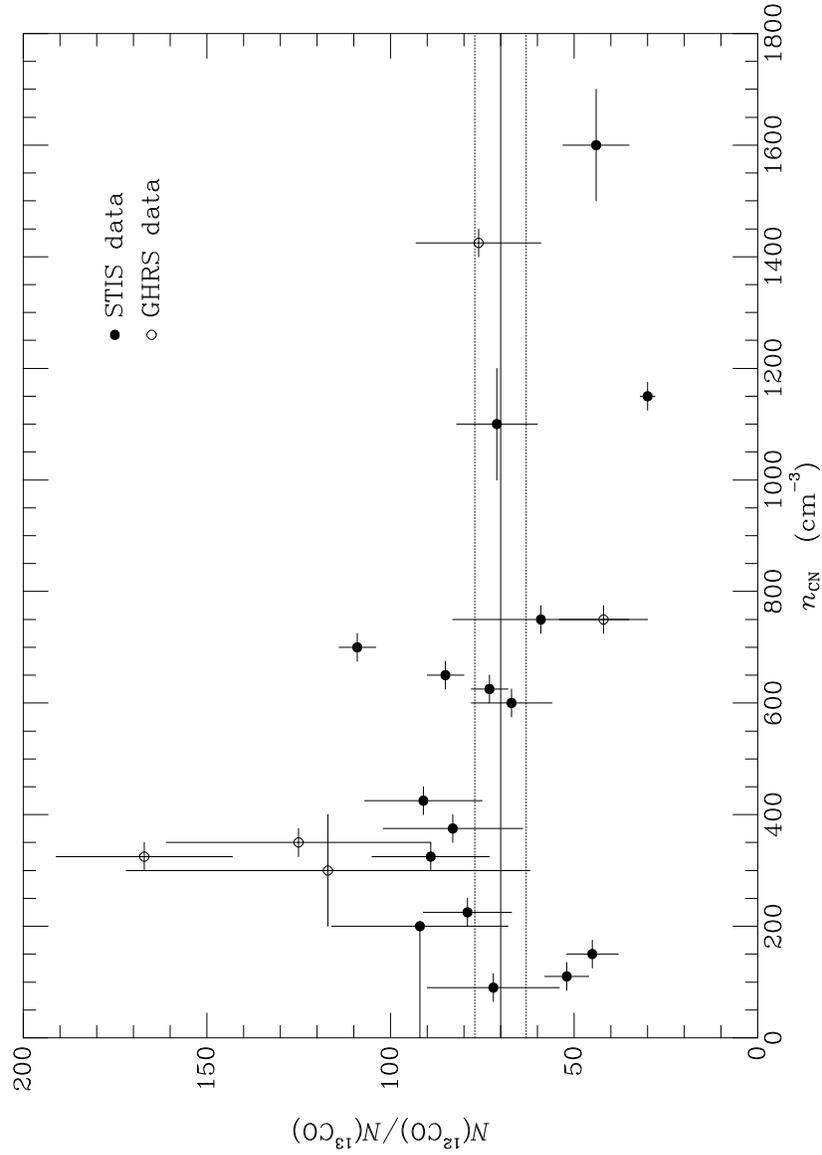}
\caption{CO isotopic ratio versus $n_{\rm CN}$.
There is no obvious correlation between the two quantities.
The solid and dashed lines denote the mean (70) and 1 standard deviation ($\pm$ 7) of
$^{12}$C/$^{13}$C in the ISM, respectively.}
\end{figure}

\end{document}